\documentclass[aps,prb,preprint,a4paper,amsmath,amssymb,showpacs]{revtex4-1}
\usepackage{graphicx}
\usepackage{booktabs,multirow}
\usepackage{hyperref}
\usepackage[normalem]{ulem}
\usepackage[svgnames]{xcolor}
\usepackage[english]{babel}

\begin{document}
\title{Finite-size scaling effect on N\'eel temperature of antiferromagnetic Cr$_2$O$_3$-(0001) films in an exchange-coupled heterostructure}
\author{Satya Prakash Pati}
\email{sppati@ecei.tohoku.ac.jp}
\author{Muftah Al-Mahdawi}
\email{mahdawi@ecei.tohoku.ac.jp}
\author{Shujun Ye}
\author{Yohei Shiokawa}
\author{Tomohiro Nozaki}
\author{Masashi Sahashi}
\affiliation{Department of Electronic Engineering, Tohoku University, Sendai 980-8579, Japan}
\date{\today}
\begin{abstract}
The scaling of antiferromagnetic ordering temperature of corundum-type chromia films have been investigated. N\'eel temperature $T_N$ was determined from the effect of perpendicular exchange-bias on the magnetization of a weakly-coupled adjacent ferromagnet. For a thick-film case, the validity of detection is confirmed by a susceptibility measurement. Detection of $T_N$ was possible down to 1-nm-thin chromia films. The scaling of ordering temperature with thickness was studied using different buffering materials, and compared with Monte-Carlo simulations. The spin-correlation length and the corresponding critical exponent were estimated, and they were consistent between experimental and simulation results. The spin-correlation length is an order of magnitude less than cubic antiferromagnets. We propose that the difference is from the change of number of exchange-coupling links in the two crystal systems.
\end{abstract}
\maketitle
\section{Introduction}
In correlated systems, the physical properties during a phase transition are altered in reduced dimensions comparable in size to a certain characteristic correlation length \cite{green_1983}. Thin films, nano-wires, and nano-particles are suitable to study confinement in one or more dimensions. The advances in fabrication of epitaxial films made them the most relevant in many technological applications. In the weakly-correlated superconductors, the correlation length is in order of tens to hundreds of nanometers \cite{merservey_1969}, thus the finite-size scaling (FSS) effects of reducing the phase-transition temperature are observed in rather thick films \cite{jin_1989}. In ferroelectrics, FSS effects appear on thinner films of a few tens of nanometers \cite{mccauley_1998}. However, in strongly-correlated systems such as ferromagnets (FMs) and antiferromagnets (AFMs) the correlation length is much shorter \cite{*[{}] [{and references therein.}] lang_2006}. Spin-correlation length measurements based on FSS in simple-cubic, body-centered cubic, and close-packed lattices were readily reported \cite{*[{}] [{and references therein.}] lang_2006}. The corresponding critical exponent estimations in these common lattices were also widely studied \cite{ritchie_1972,*[{}] [{ and references therein.}] leguillou_1980,leguillou_1985,ferer_1986,chen_1993}. However, studies on corundum-type magnetic materials are only a few \cite{he_2012}. In the report by He \emph{et al.}~\cite{he_2012}, the blocking temperature of exchange-bias data were considered rather than N\'eel temperatures. Monte-Carlo (MC) simulation studies of corundum-type Cr$_2$O$_3$ were reported before \cite{murtazaev_1999,*murtazaev_1999-1,kota_2014-1}. However, estimations of spin-correlation length and the corresponding critical exponent $\nu$ were not investigated.

The research on Cr$_2$O$_3$ has gained a renewed interest for exploration of voltage-controlled magnetic states near to room temperature \cite{borisov_2005,he_2010,ashida_2014,ashida_2015,toyoki_2015,toyoki_2015-1}. These findings opened a pathway to utilize chromia in voltage-controlled spintronic devices \cite{borisov_2005,belashchenko_2016} including hard-disk-drive media \cite{matsuzaki_2014,shibata_2015}. For applications requiring a low switching voltage, it is required to fabricate high-quality continuous ultrathin chromia films while retaining the magnetic properties. However, the thermal stability  and the operating temperature of the device may decrease by using ultrathin films, due to the FSS effect on N\'eel temperature and the low AFM anisotropy. Recently, there were some efforts to enhance the thermal stability of chromia. The effect of lattice strain induced by lattice mismatch on $T_N$ was demonstrated theoretically \cite{kota_2013}, as well as experimentally \cite{pati_2015}. Also, an enhancement of $T_N$ by boron doping in the anion sites of Cr$_2$O$_3$ was also predicted theoretically \cite{mu_2013}, and confirmed experimentally \cite{street_2014}. Another approach is the spin-correlation effect, where the length of spin correlation increases more than twice the bulk value when Cr$_2$O$_3$ is laminated with Fe$_2$O$_3$ with an oxygen-divided interface\cite{kota_2014-1}. It was reported that a strong exchange-coupling at the interface between two AFMs having different $T_N$'s and AFM anisotropies can enhance either one when the thickness reaches the spin-correlation length, as exemplified in a CoO/NiO bilayer system \cite{carey_1993}. However, an experimental determination of the spin-correlation length in Cr$_2$O$_3$ has not been reported.

FSS observations on AFMs are a challenge because of the diminishing stray magnetization. The shift in the N\'eel temperature $T_N$ in ultra-thin AFM films was detected by \emph{ac} susceptibility \cite{ambrose_1996}, neutron diffraction \cite{zaag_2000}, specific heat \cite{abarra_1996,molina-ruiz_2011}, x-ray magnetic linear dichroism \cite{park_2013}, and spin-current absorption \cite{frangou_2016}. In this report we demonstrate a simpler method to detect $T_N$ in ultrathin AFMs. By detecting the change of the equilibration angle of the magnetization of an adjacent FM layer, the onset of AFM ordering and $T_N$ can be inferred. After introducing this detection method in section \ref{sec:TN_detect}, we used it to study the shift in $T_N$ of Cr$_2$O$_3$ films with thicknesses down to 1 nm in section \ref{sec:fss}. Additionally, we compared the experimental results with MC simulations.

\section{Experimental details}
\begin{table}
\caption{\label{tab:lattice}The inplane lattice constants of a 20-nm Cr$_2$O$_3$ layer over different buffer layers \cite{pati_2015}, and for the simulated crystal. Also, the deduced values from section \ref{sec:fss} of N\'eel temperature $T_N^\infty$, correlation length $\xi_0$, and shift exponent $\lambda$ are included.}
\begin{ruledtabular}
\begin{tabular}{lcccc}
Buffer layer&$a$ [\AA]&$T_N^\infty$ [K]&$\xi_0$ [nm]&$\lambda$\\
\hline
$\alpha$-Fe$_2$O$_3$\footnotemark[1]&5.04\footnotemark[2]&266&\multirow{3}{*}{0.57(6)}&\multirow{3}{*}{1.34(7)}\\
$\alpha$-Ir-Fe$_2$O$_3$&5.02\footnotemark[2]&281&&\\
Pt\footnotemark[1]&4.98&297&&\\
\hline
Simulation 1&4.95&300&0.20(2)&1.37(2)\\
Simulation 2&4.95&306&0.24(2)&1.26(1)\\
\end{tabular}
\end{ruledtabular}
\footnotetext[1]{From Ref.~\onlinecite{pati_2015}.}
\footnotetext[2]{Assuming that Cr$_2$O$_3$ films has a pseudomorphic growth over Fe$_2$O$_3$ \cite{chambers_2000,pati_2015}.}
\end{table}

Heterostructures of Cr$_2$O$_3$ ($t_\mathrm{Cr2O3}$)/Ru ($t_\mathrm{Ru}$)/Co (1)/Pt (5) were grown over different buffer layers on \emph{c}-Al$_2$O$_3$ substrates. The numbers in parentheses represent thicknesses in nanometer. The buffer layers were Pt (25), $\alpha$-Fe$_2$O$_3$ (20) , and Ir-doped $\alpha$-Fe$_2$O$_3$ (20). The lattice mismatch with the different buffers was used to control the inplane lattice strain in Cr$_2$O$_3$ and hence $T_N$ \cite{kota_2013,pati_2015}. The buffers had different spin structures, namely non-magnetic for Pt, an in-plane spin orientation for Fe$_2$O$_3$ \cite{shimomura_2015}, and an out-of-plane spin orientation for Ir-Fe$_2$O$_3$ \cite{shimomura_2015,mitsui_2016}. It was predicted that Cr$_2$O$_3$ would have an increased spin correlation length at the interface in the bilayer of Fe$_2$O$_3$/Cr$_2$O$_3$ \cite{kota_2014}. Therefore, the investigation of correlation length over buffers with different magnetic structure is needed to explore such effects. The oxide layers were deposited by reactive radio-frequency magnetron sputtering in a mixed atmosphere of argon and oxygen from metal Cr, Fe, and Ir$_{0.1}$-Fe$_{99.9}$ targets. The (Ar, O$_2$) gas flow in sccm was fixed at (8.0, 2.0) for both of Fe$_2$O$_3$ and Ir-Fe$_2$O$_3$ growth, and (9.0, 0.85) during Cr$_2$O$_3$ growth. All of the buffer layers and Cr$_2$O$_3$ layers were grown at 773 K. The other metal layers were grown at 423 K. All of the metal layers were deposited by direct-current sputtering. Deposition rates were determined from the rate calibration and for Pt buffer, Fe$_2$O$_3$, Cr$_2$O$_3$, Ru, and Co they were 5.45, 0.17, 0.26, 1.44, and 2.86 nm/min within 3\% error, respectively.

\begin{figure}
   \caption{The height distribution of a 1.5-nm Cr$_2$O$_3$ surface grown over an Ir-Fe$_2$O$_3$ buffer. The fitting (red line) to a log-normal distribution gives a shape parameter $s$ of 0.036. The inset shows the corresponding topography scan.}
	\label{fig:roughness}
	\includegraphics[height=0.3\textheight]{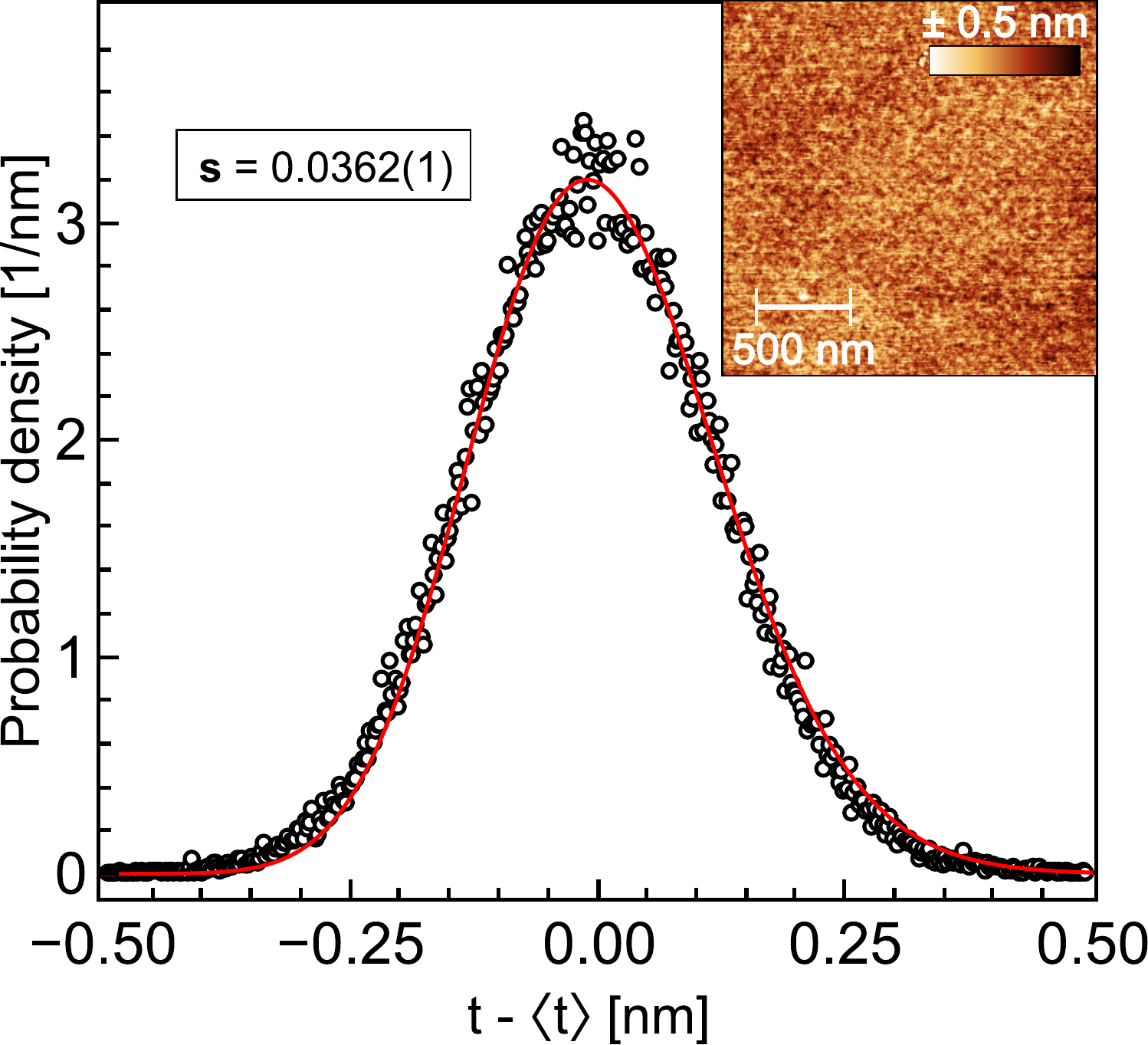}
\end{figure}

The conditions for epitaxial growth were chosen to minimize the surface roughness of each layer. In the Fe$_2$O$_3$ and Ir-Fe$_2$O$_3$ buffers case as an example, atomic-force microscopy showed a roughness average $R_a$ of less than 0.1 nm for both of the buffers and the respectively grown Cr$_2$O$_3$ layer. The surface-height histogram of a 1.5-nm Cr$_2$O$_3$ layer deposited over an Ir-Fe$_2$O$_3$ buffer is shown in Fig.~\ref{fig:roughness}, with the surface topography in the inset. {The surface is flat with $R_a = 0.09$ nm. We assume that the thickness follows a log-normal distribution of the following form:}

\begin{equation}\label{eq:P_T}
P_T(t,t_n,s) = \frac{1}{t_\mathrm{Cr2O3} s \sqrt{2 \pi}} \exp \left( \frac{\ln (t_\mathrm{Cr2O3} / t_n)}{\sqrt{2} s} \right)^2,
\end{equation}
where $t_\mathrm{Cr2O3}$ is the local thickness, $t_n$ is the median thickness, and $s$ is the shape parameter. {A fitting around the average thickness $\langle t_\mathrm{Cr2O3} \rangle$ gave $s = 0.036$. The previous assumption of using Eq.~\ref{eq:P_T} to describe the film thickness is solely based on the experimental observation. Eq.~\ref{eq:P_T} is used in section \ref{sec:fss} to estimate the error introduced by ignoring roughness.}

Analysis by X-ray diffraction \cite{pati_2015,shimomura_2016} and transmission-electron microscopy \cite{shimomura_2016} indicated the epitaxial growth, the flat sharp interfaces, and the control of lattice strain in the Cr$_2$O$_3$ layer over different buffers. The detailed structural studies are presented elsewhere \cite{pati_2015,shimomura_2016}. Table \ref{tab:lattice} summarizes the inplane lattice parameters of a 20-nm Cr$_2$O$_3$ layer over the different buffers, in addition to the lattice parameters used for the simulation mentioned afterwards. The characterization of magnetic properties was done by a commercial magnetometer based on a superconducting quantum interference device. The magnetometer was set up to measure the out-of-plane component of magnetization, and the magnetic field was applied in the out-of-plane direction. Additionally, we compared experimental results with Monte-Carlo simulations conducted using \textsc{Vampire} atomistic simulation package \cite{evans_2014}, with a Heisenberg's spin-Hamiltonian formalism.
 
\section{N\'eel temperature detection} \label{sec:TN_detect}
\begin{figure}
    \caption{(a) A schematic of $T_N$ detection method. The anisotropy of Co layer is optimized such that the ordering of Cr$_2$O$_3$ spins changes the tilting of Co magnetization from inplane above $T_N$ to an out-of-plane direction below $T_N$. A Ru metallic spacer layer is used to tune the exchange coupling energy $J_K$ and interfacial perpendicular anisotropy. The definitions of magnetic field $H$ and magnetization equilibration angle $\theta_0$ are indicated. (b) Dependence of the total exchange coupling energy $J_K$ on Ru spacer thickness $t_\mathrm{Ru}$ found from magnetization hysteresis loops at 100 K. $J_K$ monotonically decreases against the thickness of Ru. (c) Dependence of low-field magnetization normalized to saturation magnetization $M_r / M_s$ on $t_\mathrm{Ru}$. At 1.25 $\leq t_\mathrm{Ru} \leq$ 1.5 nm, a large change in $M_r/M_s$ is found above and below $T_N$. (d) The change of $M$-$H$ loops with changing $t_\mathrm{Ru}$ is shown. For thin Ru ($t_\mathrm{Ru} \leq$ 1.0), a dominant perpendicular anisotropy remains above and below $T_N$. At an intermediate thickness of 1.25--1.50 nm, there is a large change of magnetization's easy direction upon crossing $T_N$. In (b),(c), and (d) a 25-nm Pt buffer was used and the solid-lines are eye-guides.}
	\label{fig:Ru_opt}
	\includegraphics[height=0.5\textheight]{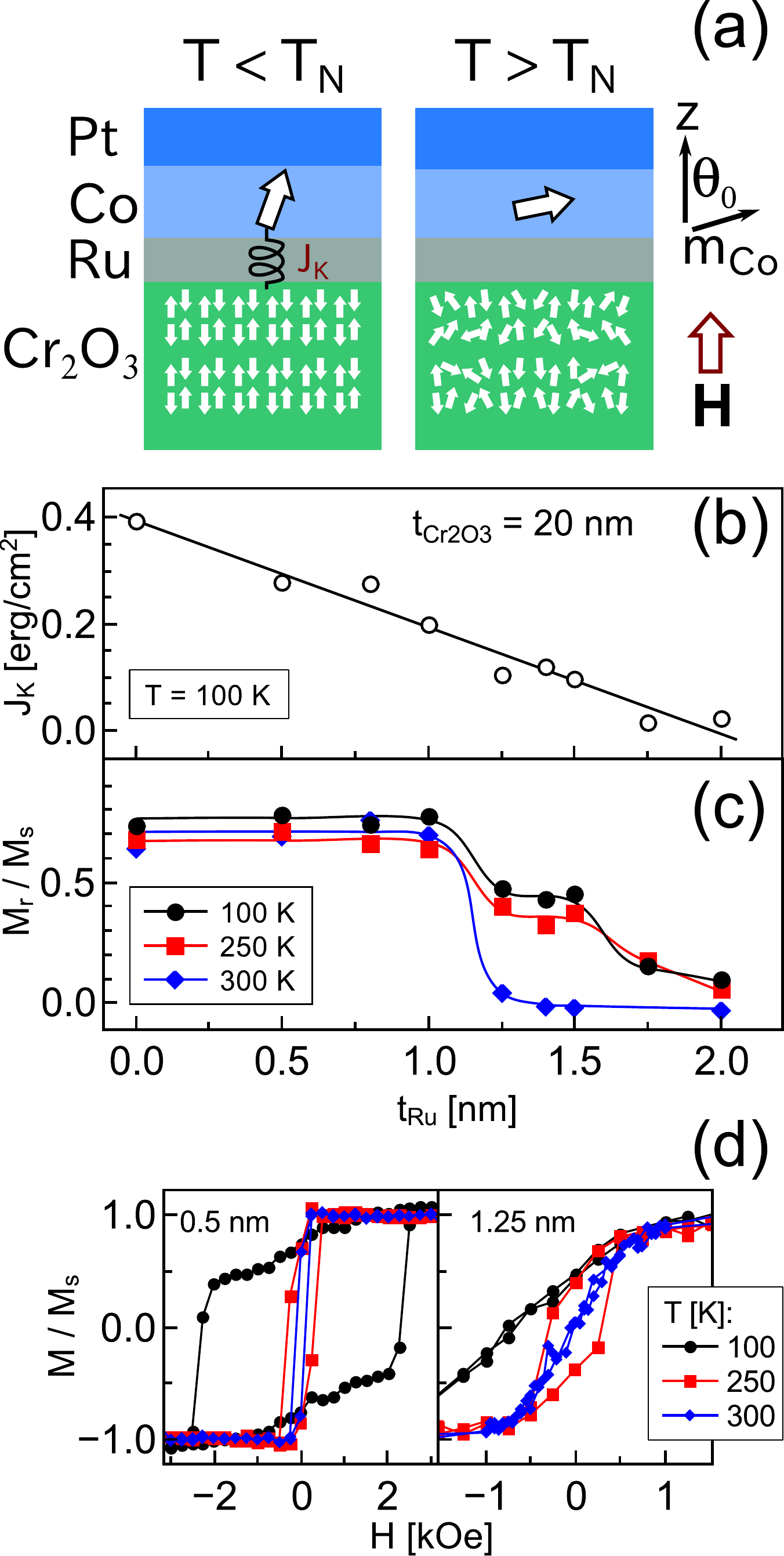}
\end{figure}

In our previous report \cite{pati_2015}, we established a detection technique of $T_N$, where a change of the magnetization in a low-field magnetization-temperature $M$-$T$ curve coincides with the enhancement of Co coercivity due to exchange coupling with AFM spins. To detect $T_N$ in ultra-thin Cr$_2$O$_3$ films ($\ll$10 nm), we optimized the anisotropy of the exchange-coupled FM layer to compensate the demagnetization field. The ordering of AFM spins at $T_N$ becomes amplified by a tilt in FM magnetization direction from in-plane to out-of-plane at a low applied magnetic field (Fig.~\ref{fig:Ru_opt}(a)). A macrospin model of the total energy per unit-area of Co magnetization, composed of demagnetization, total interfacial anisotropy, Zeeman, and exchange coupling energies can be written as follows:

\begin{align}\label{eq:WCo}
W_\mathrm{Co} &= \left(2\pi M_s^2 t_\mathrm{Co} - J_i\right) \cos^2\theta - \left(H M_s t_\mathrm{Co} + J_K \right) \cos\theta \nonumber \\
&\equiv K_\mathrm{eff} \cos^2\theta - K_{H} \cos \theta,
\end{align}
where $W_\mathrm{Co}$, $\theta$, $t_\mathrm{Co}$, $M_s$, $J_i$, $J_K$ are the total areal energy density, the magnetization angle from perpendicular direction, Co thickness, Co saturation magnetization, interfacial anisotropy energy density of the top and bottom interfaces, and exchange coupling energy with Cr$_2$O$_3$, respectively. The effective uniaxial and unidirectional anisotropies are represented by $K_\mathrm{eff}$ and $K_H$, respectively. A positive (negative) $K_\mathrm{eff}$ corresponds to an inplane (out-of-plane) easy direction of Co's magnetization. The exchange coupling energy $J_K$ is considered as an effective value representing the average exchange coupling energy through Cr$_2$O$_3$/Co interface, which is determined experimentally. Such a simplified model can be used due to the simple collinear alignment of Cr$_2$O$_3$ and Co spins, and the dominance of uncompensated surface spins at Cr$_2$O$_3$ surface \cite{belashchenko_2010}. The normalized perpendicular component of Co magnetization $m_z$ is found from the equilibration angle $\theta_0$ at which the energy is minimized with a stable solution.  The relevant solution that has a varying $m_z$ is:

\begin{equation}\label{eq:dWdth}
m_z = \cos\theta_0 = \frac{K_H}{2 K_\mathrm{eff}}, \text{where} \left| \frac{K_H}{2 K_\mathrm{eff}} \right| \leq 1.
\end{equation}
The term with the strongest temperature-dependence is $J_K$, which is proportional to the average order parameter of Cr$_2$O$_3$. The change of Co saturation magnetization in the temperature range of measurement is negligible. Therefore, the temperature dependence of $m_z$ at a fixed low field $M_r(T)$ and the accompanying change of slope $dM_r/dT$ are related to the ordering of Cr$_2$O$_3$ spins at $T_N$.
In order to obtain a large change of equilibrium angle from in-plane above $T_N$ to out-of-plane tilting below $T_N$, $K_\mathrm{eff}$ should be $0 < K_\mathrm{eff}\lessapprox J_K/2$.

To tune the interfacial anisotropy and the exchange-coupling energies, we used a Ru-metal spacer. {We found that Co has an inplane interface anisotropy with Ru, which gives another free parameter for a fine control of Co total anisotropy.} In this report, we optimized the thickness of Ru spacer to allow for the detection of $T_N$ down to $t_\mathrm{Cr2O3}$ = 1 nm. We varied the thickness of Ru $t_\mathrm{Ru}$ in the stack: Pt (25)/Cr$_2$O$_3$ (20)/Ru ($t_\mathrm{Ru}$)/Co (1)/Pt (5). Figure \ref{fig:Ru_opt}(b) shows the effect of Ru insertion on decreasing the total exchange-coupling energy $J_K$ between Co and Cr$_2$O$_3$. For a weak FM/AFM coupling compared to AFM anisotropy, $J_K$ manifests as an exchange-bias field $H_\mathrm{eb}$. In the strong coupling case, an increase of FM coercivity $\Delta H_C$ over a base value is observed. Thus, $J_K$ was determined from $(H_\mathrm{eb} + \Delta H_C)/(M_s t_\mathrm{Co})$ \cite{pati_2015}, where the experimental values of $M_s t_\mathrm{Co}$ were used \cite{shimomura_2016}. For $t_\mathrm{Ru}$ $<$ 1 nm, high squareness remained above and below $T_N$ of $\approx$290 K, and the change of $M_r$ was small (Fig.~\ref{fig:Ru_opt}(c)). At intermediate thicknesses of 1.25--1.5 nm, a large change of $M_r$ above and below $T_N$ was found. At more than 1.8 nm of Ru, the exchange coupling was diminished, and detection of $T_N$ was not feasible. Examples of magnetization hysteresis loops at strong and intermediate couplings are shown in Fig.~\ref{fig:Ru_opt}(d). We fixed $t_\mathrm{Ru}$ at 1.25 nm for all subsequent experiments. At this thickness the optimized values of $J_K$ and $K_\mathrm{eff}$ were obtained at 0.10--0.12 and  $+$0.05 erg/cm$^2$, respectively. Hence, the condition of $0<K_\mathrm{eff}\lessapprox J_K/2$ is fulfilled.

\begin{figure}
    \caption{Comparison between low-field and high-field $M$-$T$ curves of Pt (25)/ Cr$_2$O$_3$ (1000)/Ru (1.25)/Co (1)/Pt (5). To remove the contribution of the substrate's diamagnetic response, the change in the total magnetization with respect to $T_N$ is plotted against temperature. The low-field $M$-$T$ curves correspond to the change in Co magnetization direction, and to the Cr$_2$O$_3$ susceptibility response $\chi H$-$T$ at high fields. Both measurements agree on a $T_N$ of $\approx$300 K.}
	\label{fig:M-T_H}
	\includegraphics[width=0.3\textwidth]{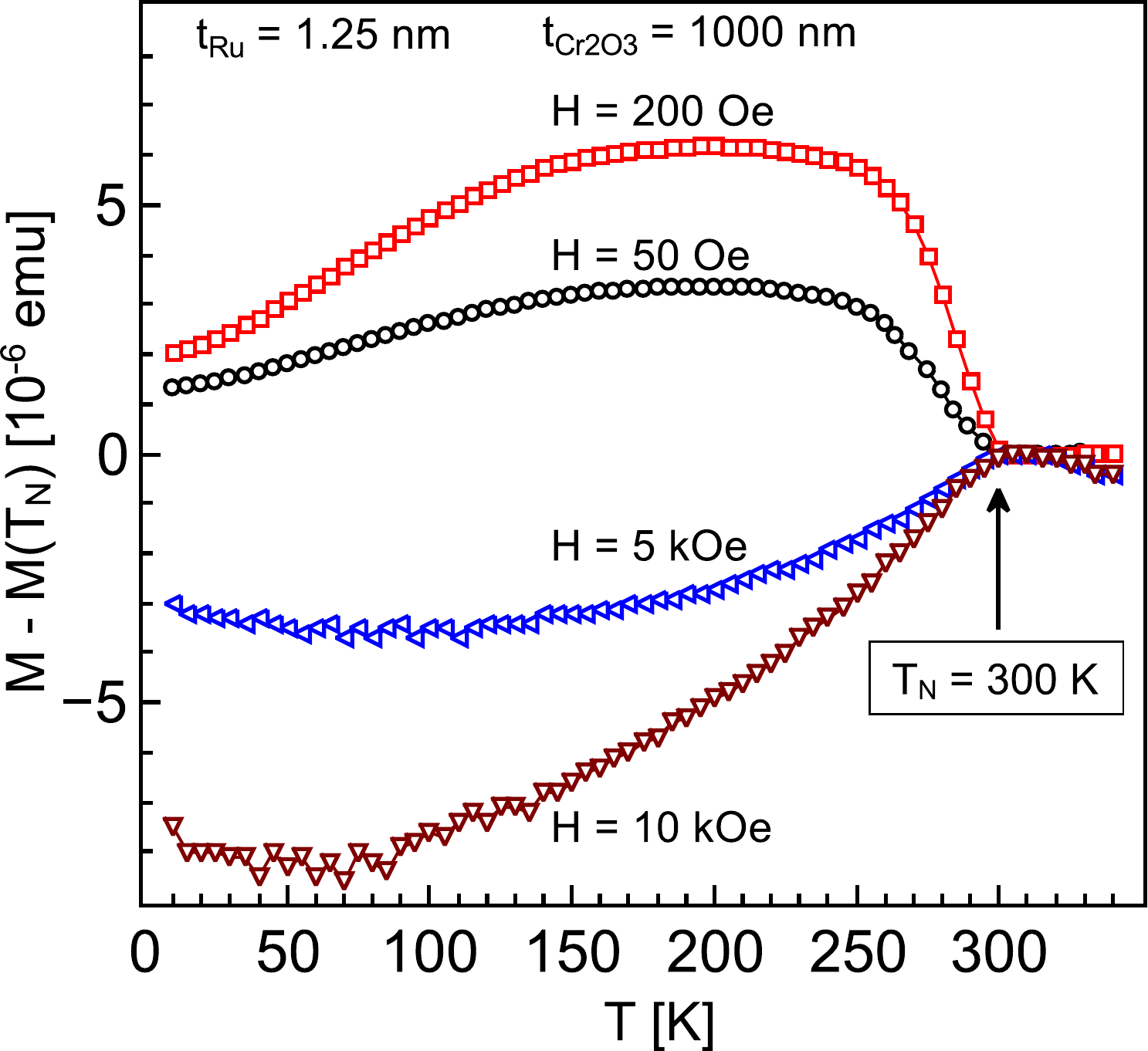}
\end{figure}

To confirm that the detected transition temperature is same as $T_N$, we compared the low-field and high-field $M$-$T$ dependencies with a rather thick Cr$_2$O$_3$ layer. The film structure was Pt (25)/Cr$_2$O$_3$ (1000)/Ru (1.25)/Co (1)/Pt (5). The total measured magnetization is composed of Co's magnetization, Cr$_2$O$_3$ antiferromangetic susceptibility response $\chi H$, and diamgnetic and paramagnetic responses of the substrate and the buffer. At low fields $<$ 500 Oe, the contribution from Cr$_2$O$_3$ bulk $\chi H$ is negligible. The features of low-field $M-T$ in Fig.~\ref{fig:M-T_H} are from the change in Co's magnetization direction at $T_N$ = 300 K as described in Eq.~\ref{eq:dWdth}. At higher fields, Co's magnetization is saturated in the out-of-plane direction, and the $M-T$ features will be from the bulk $\chi H$ of Cr$_2$O$_3$. The shape of high-field $M-T$ in Fig.~\ref{fig:M-T_H} is the same as the susceptibility parallel to [0001] growth direction of Cr$_2$O$_3$, and the cusp at 300 K corresponds to $T_N$ \cite{mcguire_1956}. The N\'eel temperature determined from both methods agreed. Thus, the low-field $M$-$T$ measurement provides an easy method to imply $T_N$ when direct detection is difficult in ultra-thin films of Cr$_2$O$_3$. In the next section, we used $M$-$T$ measurements to study the shift of $T_N$ in ultra-thin films.

The change of magnetization amplitude in the low-field $M-T$ curves is larger for $H$ = 200 Oe compared to 50 Oe (Fig.~\ref{fig:M-T_H}). The reason is not directly visible from Eq.~\ref{eq:dWdth}. The interface anisotropy $J_i$ can be decomposed into two parts $J_i = J_{i0} + \delta J_i(T)$, where $J_{i0}$ is the larger part resulting from the interface anisotropy of Co with Ru and Pt and it is weakly dependent on temperature, and $\delta J_i(T)$ is small but with a large temperature dependence, and it corresponds to a perpendicular anisotropy due to exchange coupling with Cr$_2$O$_3$. Assuming that $\delta J_i(T) \ll J_{i0}$, then Eq.~\ref{eq:dWdth} can be approximated to:

\begin{align}
\Delta m_z(T) &= m_z(T) - m_z(T>T_N) \nonumber\\
&\approx \frac{1}{2} \left( \frac{J_K(T)}{2 \pi M_s^2 t_\mathrm{Co} - J_{i0}} + \frac{\delta J_i(T) H M_s t_\mathrm{Co}}{(2 \pi M_s^2 t_\mathrm{Co} - J_{i0})^2} \right).
\end{align}
Hence, the presence of a weak perpendicular exchange anisotropy results in a larger $\Delta m_z$ for a larger $H$.

As an additional consideration, it is possible to ignore the effects of the exchange-coupling field on shifting $T_N$. The exchange-coupling field can have the same effects on AFM ordering as an applied magnetic field. It was reported that $T_N$ decreases with an external magnetic field \cite{iyama_2013}. However, the decrease is negligible on an order of 5--6 mK/kOe by application of a high magnetic field up to 90 kOe. Contrarily, in the case of a strong exchange coupling with Co without a spacer layer, the ordering of Cr$_2$O$_3$ interface layer was reported above $T_N$ \cite{shiratsuchi_2015}. In the present report, the exchange coupling is weak through the Ru spacer for $t_\mathrm{Ru} > 1.0$ nm. If there is an effect from this weak exchange coupling with Co on $T_N$, then $T_N$ will be significantly dependent on $t_\mathrm{Ru}$. No such a dependence was found.
 
\section{Finite-size scaling and correlatation length}\label{sec:fss}
Finite-size scaling effects start to be observed when one of the system's dimensions becomes comparable to the characteristic length scale, which is the spin-spin correlation length \cite{barber_1983,privman_1990}. This is mostly pronounced for thin films, nano-wires, and nano-particles. The variation of $T_N$ with thickness is expected to follow the finite-size scaling relation of\cite{fisher_1972,binder_1974,barber_1983,ambrose_1996}:

\begin{equation}\label{eq:scaling}
\Delta T_N (t) = \frac{T_N^\infty-T_N(t)}{T_N^\infty} = \left( \frac{t}{\xi_0} \right)^{-\lambda},
\end{equation}
where $\Delta T_N$ is the normalized shift in $T_N$, $T_N^\infty$ is the N\'eel temperature in the bulk, $T_N(t)$ is the shifted N\'eel temperature of the film with a finite thickness $t$, $\xi_0$ is the spin-spin correlation length at zero temperature, and $\lambda$ is the shift exponent related to the critical exponent ($\nu=1/\lambda$) governing the temperature dependence of the correlation length:

\begin{equation} \label{eq:xi-T}
\xi(T) = \xi_0 \vert 1-T/T_N \vert^{-\nu}.
\end{equation}
Bulk $T_N$ of Cr$_2$O$_3$ can be decreased (increased) by expanding (shrinking) the in-plane lattice spacing as predicted theoretically\cite{kota_2013}, and subsequently confirmed experimentally \cite{pati_2015}. However, a study on the effect of changing bulk $T_N$ on spin-spin correlation in ultra-thin chromia is still lacking.

\begin{figure}
    \caption{(a) $M$-$T$ curves under $H$ = 50 Oe for Fe$_2$O$_3$-buffered Cr$_2$O$_3$ ($t_\mathrm{Cr2O3}$)/Ru (1.25)/Co (1)/Pt (5), where $t_\mathrm{Cr2O3}$ = 20, 10, 5, 3, and 1 nm. (b) The shift of $T_N$ with $t_\mathrm{Cr2O3}$ is shown for different buffers, in addition to Monte-Carlo simulations. (c) A log-log plot of normalized $T_N$ versus $t_\mathrm{Cr2O3}$. Solid lines are fittings to Eq.~\ref{eq:scaling}. The characteristic spin-correlation length $\xi_0$ is indicated by an extrapolation to the vanishing point of $T_N$. Shift exponent $\lambda$ is determined from the slope. Both $\xi_0$ and $\lambda$ are in a reasonable agreement among experiments and simulation.}
	\label{fig:TN-scaling}
	\includegraphics[height=0.5\textheight]{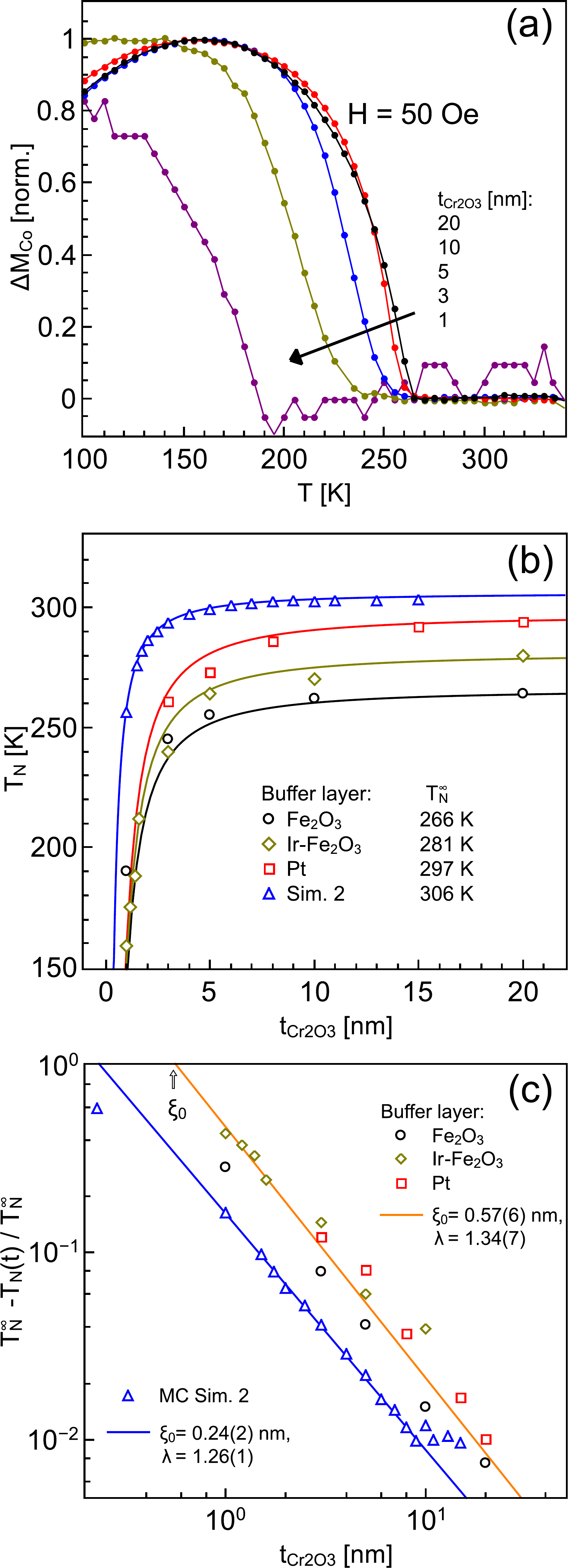}
\end{figure}

To study the effects of FSS on $T_N$ of Cr$_2$O$_3$, $t_\mathrm{Cr2O3}$ was varied from 20 nm down to 1 nm. Figure \ref{fig:TN-scaling}(a) shows M-T curves measured at a low out-of-plane field of 50 Oe of samples with a Fe$_2$O$_3$ buffer. A decrease of $T_N$ with decreasing $t_\mathrm{Cr2O3}$ is found. AFM ordering was still present down to $t_\mathrm{Cr2O3}$ = 1 nm, with $T_N$ = 195 K. By changing the buffer layer to other buffers, different $T_N^\infty$ were found due to the change in Cr$_2$O$_3$ lattice constant (table \ref{tab:lattice}). However, the reduction of $T_N$ with reducing $t_\mathrm{Cr2O3}$ was similar between buffers (Fig.~\ref{fig:TN-scaling}(b)).
Because the change in $T_N$ normalized by $T_\mathrm{N}^\infty$ did not show a significant dependence on the buffer choice, all of the experimental data of the different buffers were fitted by a single fitting to Eq.~\ref{eq:scaling} on a log-log scale (Fig.~\ref{fig:TN-scaling}(c)). The estimated values of $\xi_0$ and $\lambda$ were 0.57(6) nm and 1.34(7), respectively (table \ref{tab:lattice}).

To confirm that $T_N$ shift is due to the FSS and hence the order estimation of $\xi_0$ and $\lambda$, we simulated the temperature dependence at the magnetic transition using \textsc{Vampire} atomistic simulation package, which is based on Monte-Carlo Metropolis algorithm solution to a classical Heisenberg's spin-Hamiltonian \cite{evans_2014}. The Hamiltonian $\mathcal{H}$ is defined as follows:

\begin{equation}
\mathcal{H} = - \sum_{i \neq j} J_{ij}  \mathbf{S}_i \cdot \mathbf{S}_j - K_u \sum_i \left( \mathbf{S}_i \cdot \mathbf{c} \right)^2,
\end{equation}
where $J_{ij}$ is the exchange interaction energy between the normalized spin vectors $\mathbf{S}_i$ and $\mathbf{S}_j$, which reside at the atomic positions $i$ and $j$. A uniaxial crystalline anisotropy $K_u$ of $2\times10^{-5}$ erg/cc represents the Cr$_2$O$_3$ anisotropy along the $\mathbf{c}$-axis \cite{foner_1963}. The exchange-interaction energies of the first- and second-nearest neighbors ($J_1$, $J_2$) are the most relevant for $T_N$ determination \cite{samuelsen_1970,murtazaev_1999,kota_2013}, and they were set to 56.4 and 25.6 meV, respectively. Both of $J_1$ and $J_2$ are due to the direct exchange interaction between Cr ions, where $J_1$ is with a single neighbor in the \emph{c}-axis direction, and $J_2$ is with other three ions in the buckled Cr ions plane. A corundum-type lattice structure was simulated, where the lattice parameters were set as \emph{a} = 4.951 \AA, and \emph{c} = 13.566 \AA \cite{finger_1980}. The Cr spin magnetic moment was set to 2.48 Bohr magnetons \cite{brown_2002}.
 
\begin{figure}
    \caption{Monte-Carlo (MC) simulation results of spin-correlation length using (a,b) simulation method 1, and (c) simulation method 2. (a) A direct calculation of correlation function $\Gamma$ between center spin $\mathbf{S(0)}$ and all other spins $\mathbf{S(r)}$ in a $5\times5\times5$-nm$^3$ geometry. The temperature-dependent correlation length $\xi$ is found from a fitting to $\Gamma = r^{-1.036} e^{-r/\xi}$. (b) Temperature dependence of $\xi$ in the cases of including or disabling periodic boundary conditions. Both cases produce the same results. Inset is a log-log plot with a linear fitting to the power law in Eq.~\ref{eq:xi-T}. (c) Thickness dependence of $T_N$ determined from the sublattice magnetization $\langle M_\mathrm{sub} \rangle$. The finite thickness results in a decrease in $T_N$ when it approaches spin-correlation length. $T_N$ values are summarized in Fig.~\ref{fig:TN-scaling}(b) and (c). The estimations of $\xi_0$ and $\lambda$ from both simulation methods are in a reasonable agreement [Figs.~\ref{fig:TN-scaling}(c) and \ref{fig:MC-sim}(b)].}
	\label{fig:MC-sim}
	\includegraphics[height=0.5\textheight]{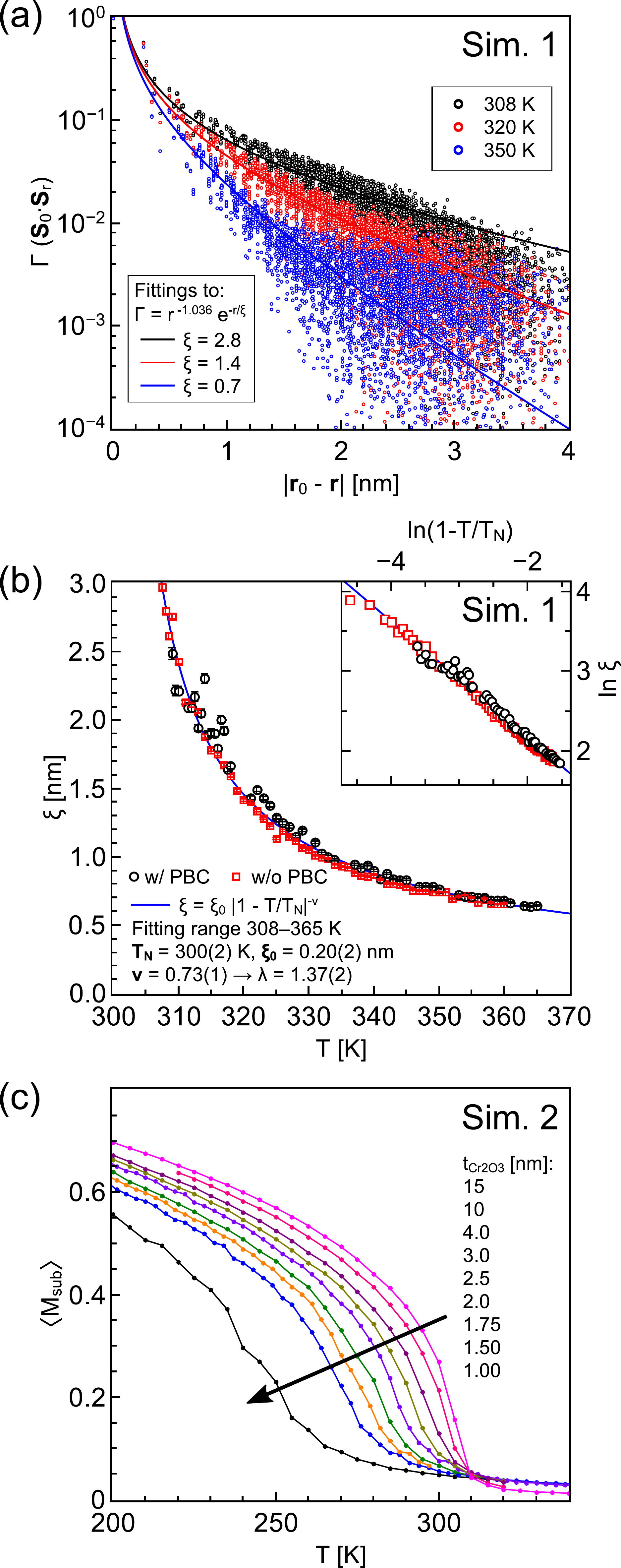}
\end{figure}

The characteristic spin-correlation length $\xi_0$ was determined by two simulation methods. In simulation 1, we directly calculated the temperature dependence of spin-correlation length and fitted it to Eq.~\ref{eq:xi-T}. The simulation geometry was a $5\times 5 \times 5$-nm$^3$ cube of 5520 Cr spins. At each temperature, $2\times 10^5$ MC steps were used for equilibration, and $5\times 10^5$ steps for time averaging. Then the correlation function $\Gamma$ between the center spin $\mathbf{S(0)}$ and all other spins $\mathbf{S(r)}$ was calculated from the recorded step-snapshots $\Gamma(\mathbf{r}) = \langle \mathbf{S(r)\cdot S(0)} \rangle$. The result was then fitted to  $\Gamma = r^{2-d-\eta} e^{-r/\xi}$, where $d=3$ is the dimensionality, and $\eta$ determines the long-range correlation near to $T_N$. The numerical value of $\eta$ is not varying considerably among 3-d models with different degrees of freedom \cite{leguillou_1980},  and a value of 0.036 was chosen. To confirm that there are no edge effects, two cases were tested with either free or periodic boundary conditions (Fig.~\ref{fig:MC-sim}(b)), and no difference was found. Also, increasing averaging steps to $8\times 10^5$ did not change results. The fitting of temperature dependence of $\xi$ gave the values of $\xi_0$ and $\nu$ at 0.20(2) nm and 0.73(1), respectively (Fig.~\ref{fig:MC-sim}(b)). It needs to be pointed that even though Cr$_2$O$_3$ should be anisotropic in $\xi_0$, we found that $\xi_0$ is almost equal along the directions parallel and perpendicular to \emph{c}-axis. The reason is likely that $J_1$ connect only a single neighbor and $J_2$ connects three neighbors, making total coupling energies similar in the parallel and perpendicular directions. Therefore, we treated $\xi_0$ as an isotropic value.

In simulation 2, we calculated $T_N$ variation with thickness in a $15\times 15 \times t_\mathrm{Cr2O3}$-nm$^3$ simulation geometry ($t_\mathrm{Cr2O3} \times 10^4$ spins). To emulate the extended films, boundary conditions were free in the thickness direction and periodic in the in-plane directions. Temperature was varied from 0 to 350 K in 5-K intervals, and averaging was taken over $2\times 10^5$ MC steps, after $10^5$ steps for equilibration. Average sublattice magnetization $\langle M_{sub} \rangle$ was same for both of spin-sublattices. For $t_\mathrm{Cr2O3} \leq 1.5$ nm, doubling the total MC steps and reducing the temperature intervals to 3 K did not affect the results. N\'eel temperature was found from a fitting of temperature dependence of $\langle M_{sub} \rangle$ to $(1-T/T_{N})^\beta$, where $\beta$ is magnetization's critical exponent. A shift of $T_N(t)$ with decreasing thickness is observed (Fig.~\ref{fig:MC-sim}(c)). Notably, the AFM ordering was maintained down to 1 nm, which is smaller than the unit cell of Cr$_2$O$_3$. This is due to the correlation length being smaller that the unit cell. Also, each sublattice is connected with other three neighbors along \emph{ab}-plane. If one of the sublattices along \emph{c}-axis is missing, the AFM order can be maintained by exchange coupling along \emph{ab}-plane. Even below 1 nm, the buckled monolayer of a 0.226-nm thickness also maintained AFM order with $T_N$ = 125 K, but it was ignored from subsequent discussion due to absence of $J_1$ coupling. The data points of $T_N(t)$ with the corresponding fitting to Eq.~\ref{eq:scaling} are shown in figures \ref{fig:TN-scaling}(b,c). The bulk $T_N$ of 306 K in units of $J_1/k_B$ is 0.468, where $k_B$ is the Boltzmann constant, which is in agreement with previous reports \cite{murtazaev_1999,*murtazaev_1999-1}. The fitted values of $\xi_0$, $\lambda$, and $\nu$ are 0.24(2) nm, 1.26(1), and 0.79(1), respectively (Fig.~\ref{fig:TN-scaling}(c)). There is a quantitative agreement between the two calculation methods. Also, the critical exponent $\nu$ is close to what is expected from the 3-d Ising, XY or Heisenberg universality models having $\nu$ in the range of 0.63--0.71 \cite{leguillou_1980,ferer_1986}.

Noting the different lattice spacings and exchange-coupling energies, there are quantitative agreements within an order of magnitude in $\xi_0$ estimations between MC calculations and experiments (Fig.~\ref{fig:TN-scaling}(c)). The difference in $\xi_0$ can be attributed to the simplifications assumed about the coupling energies in the MC calculations. Namely, the far-ranged exchange-coupling of third- to fifth-nearest neighbors and the strain field should increase $\xi_0$. Both of the experimental and simulation values of the shift exponent $\lambda$ values agree reasonably with 3-d universality models. So we can conclude that the observed reduction in transition temperature is due to the FSS. 

Concerning the effect of surface roughness on the extraction of FSS parameters, the Cr$_2$O$_3$ layer can be considered as composed of smaller areas with slowly-varying thicknesses. The local thickness distribution can be represented by Eq.~\ref{eq:P_T}. The exchange-coupling energy between Co and Cr$_2$O$_3$ layers is weaker than the exchange stiffness of Co. Therefore, the response of Co's magnetization is the average of AFM ordering in Cr$_2$O$_3$, and the measured $T_N$ is the average of the whole Cr$_2$O$_3$ layer. The average normalized shift of $T_N$ ($\langle \Delta T_N \rangle$) can be found as follows:

\begin{align}\label{eq:avg_DTN}
\langle \Delta T_N \rangle &= \int_0^\infty \Delta T_N P_T dt \nonumber \\
&=\left( \frac{\langle t_\mathrm{Cr2O3} \rangle}{\xi_0} \right)^{-\lambda} \exp \left( \frac{1}{2} s^2 (\lambda+1) \right),
\end{align}
where $\langle t_\mathrm{Cr2O3} \rangle = \exp ( \ln t_n + s^2/2 )$ is the average thickness. As a first-order approximation for small $s^2$, the deviation caused by using an average global thickness and neglecting roughness is on order of $s^2/2 (\lambda+1)$. This results in $<$5\% error for $s=0.2$. Therefore, the FSS relation is relatively insensitive to size distribution, {as it was shown in  Ref.~\cite{molina-ruiz_2011}. However, we did not take into account the effects of magnetically-dead interfacial layers and interdiffusion. We assumed them to be minimal in the present study, but such an assumption is not always safe, as shown by Ref.~\cite{tang_2003}.}

\begin{figure}
    \caption{Plane projections of (a) corundum and (b) rock-salt crystal structures on ($11\bar{2}0$) and ($100$) planes, respectively. Differences in the coordination number of exchange interactions result in a different correlation length for each crystal system. It can be seen by counting the number of indirect connections between far neighbors, \emph{e.g.}~spins marked by numbers 1 and 2. The rock-salt structure has more connections in comparison to corundum-type structure.}
	\label{fig:corun-fcc}
	\includegraphics[height=0.3\textheight]{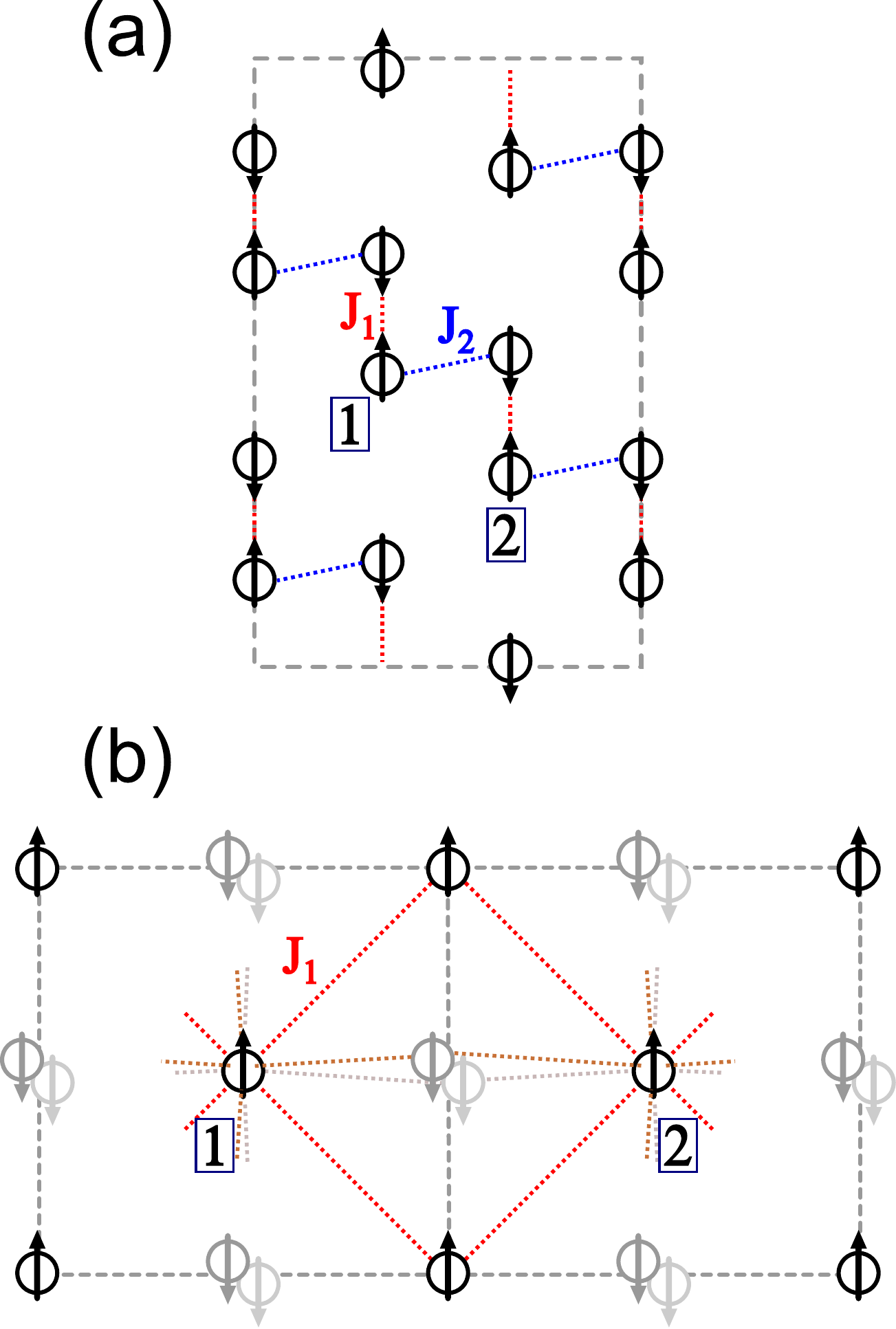}
\end{figure}

The spin-correlation length of 0.2--0.6 in Cr$_2$O$_3$ is much shorter than what was reported for CoO and NiO of 1.0--2.1 nm and 1.4 nm, respectively \cite{ambrose_1996,molina-ruiz_2011,abarra_1996,lang_2006}. We attribute this difference to the difference in the coordination number of exchange interactions. In the following discussion, the interaction coordination number without regard to non-magnetic ions is considered. Also, we define the nearest-neighbor degree $n$NN as the number of intermediate magnetic ions that relay the exchange coupling, so that the first-nearest-neighbors 1NN are the spins with direct coupling regardless of being either of superexchange or direct-exchange type. In that sense, in the Cr$_2$O$_3$ structure, both $J_1$ and $J_2$ are connecting 1NN's. In a simplistic model, the number of paths between far neighbors ($\geq$ 2NN) in a corundum-type Cr$_2$O$_3$ crystal are limited to one, \emph{e.g.}~spins numbered 1 and 2 in Fig.~\ref{fig:corun-fcc}(a). On the other hand, in the rock-salt structure with only 1NN interactions, there are at least 4 paths of coupling between 2NN's, which are marked as 1 and 2 in Fig.~\ref{fig:corun-fcc}(b). Also, the number of paths does not decay rapidly with distance. This can explain qualitatively the difference between correlation lengths in close-packed crystals and corundum-type.

In the discussion above, we only considered Cr$_2$O$_3$ with $J_1$ and $J_2$ interactions. Adding more far-ranged exchange interactions can increase the correlation length. Kota \emph{et al.}~reported a doubling of correlation length in Cr$_2$O$_3$ at the interface with an Fe$_2$O$_3$ layer \cite{kota_2014}. This is due to additional superexchange-type interactions with a longer range at the oxygen-divided interface. However, we did not find an effect of Fe$_2$O$_3$ or Ir-Fe$_2$O$_3$ buffer layers on the relative $T_N$ shift of Cr$_2$O$_3$ (Fig.~\ref{fig:TN-scaling}(c)). In order to minimize the surface charge, the corundum-type crystals prefer to terminate in the bottom layer of the buckled metal layer (ion No.~2 in Fig.~\ref{fig:corun-fcc}(a)) \cite{wang_2000}. Therefore, it is likely that the interface of Fe$_2$O$_3$/Cr$_2$O$_3$ is a metal-split one, which is not different from a stand-alone Cr$_2$O$_3$ crystal \cite{kota_2014}. If an oxygen-divided interface can be realized, the correlation length can be increased by the effect of Fe$_2$O$_3$ buffer layer.

\section{Conclusion}
We presented a study on the effect of finite-size scaling on the antiferromagnetic order of corundum-type Cr$_2$O$_3$ ultra-thin films. The films were epitaxially grown by reactive sputtering, and were 1--20 nm in thickness. The N\'eel temperature $T_N$ was determined by a relatively-easy method of measuring the effect of an optimized exchange-coupling on a proximate ferromagnetic layer. By controlling the lattice spacings of Cr$_2$O$_3$ films, different bulk N\'eel temperature values $T_N^\infty$ could be achieved on different buffer layers. For each buffer layer, $T_N$ monotonically decreased when the film thickness was decreased, in accordance with finite-size scaling. The spin-correlation length $\xi_0$ and the shift exponent $\lambda$ obtained from the experimental results did not show a significant dependence on $T_N^\infty$ and the choice of buffer layer. Monte-Carlo simulations of the spin-spin correlation function and the finite-size effects also agreed reasonably with the experimental results. Moreover, the shift exponents $\lambda$ were close with the expectations from three-dimensional universalities. We found that $\xi_0$ of Cr$_2$O$_3$ was much smaller than what was reported for CoO, NiO, and other close-packed crystals. We attribute this change to the difference between corundum-type and rock-salt-type crystals in the number of exchange-coupling paths between far neighbors. The understanding the critical behavior of ultra-thin Cr$_2$O$_3$ films should pave the way for more work on realizing ultra-thin magnetoelectric storage media having a high N\'eel temperature.

\begin{acknowledgments}
The authors thank Dr.~Richard Evans for his comments and the \textsc{Vampire} community for their support. This work was partly funded by ImPACT Program of Council for Science, Technology and Innovation (Cabinet Office, Japan Government).
\end{acknowledgments}

\clearpage
\newpage
\bibliographystyle{apsrev4-1}

\bibliography{refs}

\begin{thebibliography}{54}%
\makeatletter
\providecommand \@ifxundefined [1]{%
 \@ifx{#1\undefined}
}%
\providecommand \@ifnum [1]{%
 \ifnum #1\expandafter \@firstoftwo
 \else \expandafter \@secondoftwo
 \fi
}%
\providecommand \@ifx [1]{%
 \ifx #1\expandafter \@firstoftwo
 \else \expandafter \@secondoftwo
 \fi
}%
\providecommand \natexlab [1]{#1}%
\providecommand \enquote  [1]{``#1''}%
\providecommand \bibnamefont  [1]{#1}%
\providecommand \bibfnamefont [1]{#1}%
\providecommand \citenamefont [1]{#1}%
\providecommand \href@noop [0]{\@secondoftwo}%
\providecommand \href [0]{\begingroup \@sanitize@url \@href}%
\providecommand \@href[1]{\@@startlink{#1}\@@href}%
\providecommand \@@href[1]{\endgroup#1\@@endlink}%
\providecommand \@sanitize@url [0]{\catcode `\\12\catcode `\$12\catcode
  `\&12\catcode `\#12\catcode `\^12\catcode `\_12\catcode `\%12\relax}%
\providecommand \@@startlink[1]{}%
\providecommand \@@endlink[0]{}%
\providecommand \url  [0]{\begingroup\@sanitize@url \@url }%
\providecommand \@url [1]{\endgroup\@href {#1}{\urlprefix }}%
\providecommand \urlprefix  [0]{URL }%
\providecommand \Eprint [0]{\href }%
\providecommand \doibase [0]{http://dx.doi.org/}%
\providecommand \selectlanguage [0]{\@gobble}%
\providecommand \bibinfo  [0]{\@secondoftwo}%
\providecommand \bibfield  [0]{\@secondoftwo}%
\providecommand \translation [1]{[#1]}%
\providecommand \BibitemOpen [0]{}%
\providecommand \bibitemStop [0]{}%
\providecommand \bibitemNoStop [0]{.\EOS\space}%
\providecommand \EOS [0]{\spacefactor3000\relax}%
\providecommand \BibitemShut  [1]{\csname bibitem#1\endcsname}%
\let\auto@bib@innerbib\@empty
\bibitem [{\citenamefont {Green}\ and\ \citenamefont
  {Lebowitz}(1983)}]{green_1983}%
  \BibitemOpen
  \bibfield  {author} {\bibinfo {author} {\bibfnamefont {M.~S.}\ \bibnamefont
  {Green}}\ and\ \bibinfo {author} {\bibfnamefont {J.~L.}\ \bibnamefont
  {Lebowitz}},\ }\href@noop {} {\emph {\bibinfo {title} {Phase transitions and
  critical phenomena}}},\ Vol.~\bibinfo {volume} {8}\ (\bibinfo  {publisher}
  {{Academic Press}},\ \bibinfo {year} {1983})\BibitemShut {NoStop}%
\bibitem [{\citenamefont {Merservey}\ and\ \citenamefont
  {Schwartz}(1969)}]{merservey_1969}%
  \BibitemOpen
  \bibfield  {author} {\bibinfo {author} {\bibfnamefont {R.}~\bibnamefont
  {Merservey}}\ and\ \bibinfo {author} {\bibfnamefont {B.~B.}\ \bibnamefont
  {Schwartz}},\ }in\ \href@noop {} {\emph {\bibinfo {booktitle}
  {Superconductivity}}},\ Vol.~\bibinfo {volume} {1},\ \bibinfo {editor}
  {edited by\ \bibinfo {editor} {\bibfnamefont {R.~D.}\ \bibnamefont {Parks}}}\
  (\bibinfo  {publisher} {{Marcel Dekker, Inc.}},\ \bibinfo {year} {1969})\ p.\
  \bibinfo {pages} {117}\BibitemShut {NoStop}%
\bibitem [{\citenamefont {Jin}\ and\ \citenamefont
  {Ketterson}(1989)}]{jin_1989}%
  \BibitemOpen
  \bibfield  {author} {\bibinfo {author} {\bibfnamefont {B.~Y.}\ \bibnamefont
  {Jin}}\ and\ \bibinfo {author} {\bibfnamefont {J.~B.}\ \bibnamefont
  {Ketterson}},\ }\href {\doibase 10.1080/00018738900101112} {\bibfield
  {journal} {\bibinfo  {journal} {Advances in Physics}\ }\textbf {\bibinfo
  {volume} {38}},\ \bibinfo {pages} {189} (\bibinfo {year} {1989})}\BibitemShut
  {NoStop}%
\bibitem [{\citenamefont {McCauley}\ \emph {et~al.}(1998)\citenamefont
  {McCauley}, \citenamefont {Newnham},\ and\ \citenamefont
  {Randall}}]{mccauley_1998}%
  \BibitemOpen
  \bibfield  {author} {\bibinfo {author} {\bibfnamefont {D.}~\bibnamefont
  {McCauley}}, \bibinfo {author} {\bibfnamefont {R.~E.}\ \bibnamefont
  {Newnham}}, \ and\ \bibinfo {author} {\bibfnamefont {C.~A.}\ \bibnamefont
  {Randall}},\ }\href {\doibase 10.1111/j.1151-2916.1998.tb02435.x} {\bibfield
  {journal} {\bibinfo  {journal} {Journal of the American Ceramic Society}\
  }\textbf {\bibinfo {volume} {81}},\ \bibinfo {pages} {979} (\bibinfo {year}
  {1998})}\BibitemShut {NoStop}%
\bibitem [{\citenamefont {Lang}\ \emph {et~al.}(2006)\citenamefont {Lang},
  \citenamefont {Zheng},\ and\ \citenamefont {Jiang}}]{lang_2006}%
  \BibitemOpen
  \bibfield  {author} {\bibinfo {author} {\bibfnamefont {X.~Y.}\ \bibnamefont
  {Lang}}, \bibinfo {author} {\bibfnamefont {W.~T.}\ \bibnamefont {Zheng}}, \
  and\ \bibinfo {author} {\bibfnamefont {Q.}~\bibnamefont {Jiang}},\ }\href
  {\doibase 10.1103/PhysRevB.73.224444} {\bibfield  {journal} {\bibinfo
  {journal} {Physical Review B}\ }\textbf {\bibinfo {volume} {73}},\ \bibinfo
  {pages} {224444} (\bibinfo {year} {2006})}\BibitemShut {NoStop}%
\bibitem [{\citenamefont {Ritchie}\ and\ \citenamefont
  {Fisher}(1972)}]{ritchie_1972}%
  \BibitemOpen
  \bibfield  {author} {\bibinfo {author} {\bibfnamefont {D.~S.}\ \bibnamefont
  {Ritchie}}\ and\ \bibinfo {author} {\bibfnamefont {M.~E.}\ \bibnamefont
  {Fisher}},\ }\href {\doibase 10.1103/PhysRevB.5.2668} {\bibfield  {journal}
  {\bibinfo  {journal} {Physical Review B}\ }\textbf {\bibinfo {volume} {5}},\
  \bibinfo {pages} {2668} (\bibinfo {year} {1972})}\BibitemShut {NoStop}%
\bibitem [{\citenamefont {{Le Guillou}}\ and\ \citenamefont
  {Zinn-Justin}(1980)}]{leguillou_1980}%
  \BibitemOpen
  \bibfield  {author} {\bibinfo {author} {\bibfnamefont {J.~C.}\ \bibnamefont
  {{Le Guillou}}}\ and\ \bibinfo {author} {\bibfnamefont {J.}~\bibnamefont
  {Zinn-Justin}},\ }\href {\doibase 10.1103/PhysRevB.21.3976} {\bibfield
  {journal} {\bibinfo  {journal} {Physical Review B}\ }\textbf {\bibinfo
  {volume} {21}},\ \bibinfo {pages} {3976} (\bibinfo {year}
  {1980})}\BibitemShut {NoStop}%
\bibitem [{\citenamefont {{Le Guillou}}\ and\ \citenamefont
  {Zinn-Justin}(1985)}]{leguillou_1985}%
  \BibitemOpen
  \bibfield  {author} {\bibinfo {author} {\bibfnamefont {J.~C.}\ \bibnamefont
  {{Le Guillou}}}\ and\ \bibinfo {author} {\bibfnamefont {J.}~\bibnamefont
  {Zinn-Justin}},\ }\href {\doibase 10.1051/jphyslet:01985004604013700}
  {\bibfield  {journal} {\bibinfo  {journal} {Journal de Physique Lettres}\
  }\textbf {\bibinfo {volume} {46}},\ \bibinfo {pages} {5} (\bibinfo {year}
  {1985})}\BibitemShut {NoStop}%
\bibitem [{\citenamefont {Ferer}\ and\ \citenamefont
  {Hamid-Aidinejad}(1986)}]{ferer_1986}%
  \BibitemOpen
  \bibfield  {author} {\bibinfo {author} {\bibfnamefont {M.}~\bibnamefont
  {Ferer}}\ and\ \bibinfo {author} {\bibfnamefont {A.}~\bibnamefont
  {Hamid-Aidinejad}},\ }\href {\doibase 10.1103/PhysRevB.34.6481} {\bibfield
  {journal} {\bibinfo  {journal} {Physical Review B}\ }\textbf {\bibinfo
  {volume} {34}},\ \bibinfo {pages} {6481} (\bibinfo {year}
  {1986})}\BibitemShut {NoStop}%
\bibitem [{\citenamefont {Chen}\ \emph {et~al.}(1993)\citenamefont {Chen},
  \citenamefont {Ferrenberg},\ and\ \citenamefont {Landau}}]{chen_1993}%
  \BibitemOpen
  \bibfield  {author} {\bibinfo {author} {\bibfnamefont {K.}~\bibnamefont
  {Chen}}, \bibinfo {author} {\bibfnamefont {A.~M.}\ \bibnamefont
  {Ferrenberg}}, \ and\ \bibinfo {author} {\bibfnamefont {D.~P.}\ \bibnamefont
  {Landau}},\ }\href {\doibase 10.1103/PhysRevB.48.3249} {\bibfield  {journal}
  {\bibinfo  {journal} {Physical Review B}\ }\textbf {\bibinfo {volume} {48}},\
  \bibinfo {pages} {3249} (\bibinfo {year} {1993})}\BibitemShut {NoStop}%
\bibitem [{\citenamefont {He}\ \emph {et~al.}(2012)\citenamefont {He},
  \citenamefont {Echtenkamp},\ and\ \citenamefont {Binek}}]{he_2012}%
  \BibitemOpen
  \bibfield  {author} {\bibinfo {author} {\bibfnamefont {X.}~\bibnamefont
  {He}}, \bibinfo {author} {\bibfnamefont {W.}~\bibnamefont {Echtenkamp}}, \
  and\ \bibinfo {author} {\bibfnamefont {C.}~\bibnamefont {Binek}},\ }\href
  {\doibase 10.1080/00150193.2012.671113} {\bibfield  {journal} {\bibinfo
  {journal} {Ferroelectrics}\ }\textbf {\bibinfo {volume} {426}},\ \bibinfo
  {pages} {81} (\bibinfo {year} {2012})}\BibitemShut {NoStop}%
\bibitem [{\citenamefont {Murtazaev}(1999)}]{murtazaev_1999}%
  \BibitemOpen
  \bibfield  {author} {\bibinfo {author} {\bibfnamefont {A.~K.}\ \bibnamefont
  {Murtazaev}},\ }\href {\doibase 10.1063/1.593747} {\bibfield  {journal}
  {\bibinfo  {journal} {Low Temperature Physics}\ }\textbf {\bibinfo {volume}
  {25}},\ \bibinfo {pages} {344} (\bibinfo {year} {1999})}\BibitemShut
  {NoStop}%
\bibitem [{\citenamefont {Murtazaev}\ \emph {et~al.}(1999)\citenamefont
  {Murtazaev}, \citenamefont {Kamilov},\ and\ \citenamefont
  {Aliev}}]{murtazaev_1999-1}%
  \BibitemOpen
  \bibfield  {author} {\bibinfo {author} {\bibfnamefont {A.~K.}\ \bibnamefont
  {Murtazaev}}, \bibinfo {author} {\bibfnamefont {I.~K.}\ \bibnamefont
  {Kamilov}}, \ and\ \bibinfo {author} {\bibfnamefont {K.~K.}\ \bibnamefont
  {Aliev}},\ }\href {\doibase 10.1016/S0304-8853(99)00168-7} {\bibfield
  {journal} {\bibinfo  {journal} {Journal of Magnetism and Magnetic Materials}\
  }\textbf {\bibinfo {volume} {204}},\ \bibinfo {pages} {151} (\bibinfo {year}
  {1999})}\BibitemShut {NoStop}%
\bibitem [{\citenamefont {Kota}\ \emph
  {et~al.}(2014{\natexlab{a}})\citenamefont {Kota}, \citenamefont {Imamura},\
  and\ \citenamefont {Sasaki}}]{kota_2014-1}%
  \BibitemOpen
  \bibfield  {author} {\bibinfo {author} {\bibfnamefont {Y.}~\bibnamefont
  {Kota}}, \bibinfo {author} {\bibfnamefont {H.}~\bibnamefont {Imamura}}, \
  and\ \bibinfo {author} {\bibfnamefont {M.}~\bibnamefont {Sasaki}},\ }\href
  {\doibase 10.1063/1.4865780} {\bibfield  {journal} {\bibinfo  {journal}
  {Journal of Applied Physics}\ }\textbf {\bibinfo {volume} {115}},\ \bibinfo
  {pages} {17D719} (\bibinfo {year} {2014}{\natexlab{a}})}\BibitemShut
  {NoStop}%
\bibitem [{\citenamefont {Borisov}\ \emph {et~al.}(2005)\citenamefont
  {Borisov}, \citenamefont {Hochstrat}, \citenamefont {Chen}, \citenamefont
  {Kleemann},\ and\ \citenamefont {Binek}}]{borisov_2005}%
  \BibitemOpen
  \bibfield  {author} {\bibinfo {author} {\bibfnamefont {P.}~\bibnamefont
  {Borisov}}, \bibinfo {author} {\bibfnamefont {A.}~\bibnamefont {Hochstrat}},
  \bibinfo {author} {\bibfnamefont {X.}~\bibnamefont {Chen}}, \bibinfo {author}
  {\bibfnamefont {W.}~\bibnamefont {Kleemann}}, \ and\ \bibinfo {author}
  {\bibfnamefont {C.}~\bibnamefont {Binek}},\ }\href {\doibase
  10.1103/PhysRevLett.94.117203} {\bibfield  {journal} {\bibinfo  {journal}
  {Physical Review Letters}\ }\textbf {\bibinfo {volume} {94}},\ \bibinfo
  {pages} {117203} (\bibinfo {year} {2005})}\BibitemShut {NoStop}%
\bibitem [{\citenamefont {He}\ \emph {et~al.}(2010)\citenamefont {He},
  \citenamefont {Wang}, \citenamefont {Wu}, \citenamefont {Caruso},
  \citenamefont {Vescovo}, \citenamefont {Belashchenko}, \citenamefont
  {Dowben},\ and\ \citenamefont {Binek}}]{he_2010}%
  \BibitemOpen
  \bibfield  {author} {\bibinfo {author} {\bibfnamefont {X.}~\bibnamefont
  {He}}, \bibinfo {author} {\bibfnamefont {Y.}~\bibnamefont {Wang}}, \bibinfo
  {author} {\bibfnamefont {N.}~\bibnamefont {Wu}}, \bibinfo {author}
  {\bibfnamefont {A.~N.}\ \bibnamefont {Caruso}}, \bibinfo {author}
  {\bibfnamefont {E.}~\bibnamefont {Vescovo}}, \bibinfo {author} {\bibfnamefont
  {K.~D.}\ \bibnamefont {Belashchenko}}, \bibinfo {author} {\bibfnamefont
  {P.~A.}\ \bibnamefont {Dowben}}, \ and\ \bibinfo {author} {\bibfnamefont
  {C.}~\bibnamefont {Binek}},\ }\href {\doibase 10.1038/nmat2785} {\bibfield
  {journal} {\bibinfo  {journal} {Nature Materials}\ }\textbf {\bibinfo
  {volume} {9}},\ \bibinfo {pages} {579} (\bibinfo {year} {2010})}\BibitemShut
  {NoStop}%
\bibitem [{\citenamefont {Ashida}\ \emph {et~al.}(2014)\citenamefont {Ashida},
  \citenamefont {Oida}, \citenamefont {Shimomura}, \citenamefont {Nozaki},
  \citenamefont {Shibata},\ and\ \citenamefont {Sahashi}}]{ashida_2014}%
  \BibitemOpen
  \bibfield  {author} {\bibinfo {author} {\bibfnamefont {T.}~\bibnamefont
  {Ashida}}, \bibinfo {author} {\bibfnamefont {M.}~\bibnamefont {Oida}},
  \bibinfo {author} {\bibfnamefont {N.}~\bibnamefont {Shimomura}}, \bibinfo
  {author} {\bibfnamefont {T.}~\bibnamefont {Nozaki}}, \bibinfo {author}
  {\bibfnamefont {T.}~\bibnamefont {Shibata}}, \ and\ \bibinfo {author}
  {\bibfnamefont {M.}~\bibnamefont {Sahashi}},\ }\href {\doibase
  10.1063/1.4871515} {\bibfield  {journal} {\bibinfo  {journal} {Applied
  Physics Letters}\ }\textbf {\bibinfo {volume} {104}},\ \bibinfo {pages}
  {152409} (\bibinfo {year} {2014})}\BibitemShut {NoStop}%
\bibitem [{\citenamefont {Ashida}\ \emph {et~al.}(2015)\citenamefont {Ashida},
  \citenamefont {Oida}, \citenamefont {Shimomura}, \citenamefont {Nozaki},
  \citenamefont {Shibata},\ and\ \citenamefont {Sahashi}}]{ashida_2015}%
  \BibitemOpen
  \bibfield  {author} {\bibinfo {author} {\bibfnamefont {T.}~\bibnamefont
  {Ashida}}, \bibinfo {author} {\bibfnamefont {M.}~\bibnamefont {Oida}},
  \bibinfo {author} {\bibfnamefont {N.}~\bibnamefont {Shimomura}}, \bibinfo
  {author} {\bibfnamefont {T.}~\bibnamefont {Nozaki}}, \bibinfo {author}
  {\bibfnamefont {T.}~\bibnamefont {Shibata}}, \ and\ \bibinfo {author}
  {\bibfnamefont {M.}~\bibnamefont {Sahashi}},\ }\href {\doibase
  10.1063/1.4916826} {\bibfield  {journal} {\bibinfo  {journal} {Applied
  Physics Letters}\ }\textbf {\bibinfo {volume} {106}},\ \bibinfo {pages}
  {132407} (\bibinfo {year} {2015})}\BibitemShut {NoStop}%
\bibitem [{\citenamefont {Toyoki}\ \emph
  {et~al.}(2015{\natexlab{a}})\citenamefont {Toyoki}, \citenamefont
  {Shiratsuchi}, \citenamefont {Kobane}, \citenamefont {Harimoto},
  \citenamefont {Onoue}, \citenamefont {Nomura},\ and\ \citenamefont
  {Nakatani}}]{toyoki_2015}%
  \BibitemOpen
  \bibfield  {author} {\bibinfo {author} {\bibfnamefont {K.}~\bibnamefont
  {Toyoki}}, \bibinfo {author} {\bibfnamefont {Y.}~\bibnamefont {Shiratsuchi}},
  \bibinfo {author} {\bibfnamefont {A.}~\bibnamefont {Kobane}}, \bibinfo
  {author} {\bibfnamefont {S.}~\bibnamefont {Harimoto}}, \bibinfo {author}
  {\bibfnamefont {S.}~\bibnamefont {Onoue}}, \bibinfo {author} {\bibfnamefont
  {H.}~\bibnamefont {Nomura}}, \ and\ \bibinfo {author} {\bibfnamefont
  {R.}~\bibnamefont {Nakatani}},\ }\href {\doibase 10.1063/1.4906322}
  {\bibfield  {journal} {\bibinfo  {journal} {Journal of Applied Physics}\
  }\textbf {\bibinfo {volume} {117}},\ \bibinfo {pages} {17D902} (\bibinfo
  {year} {2015}{\natexlab{a}})}\BibitemShut {NoStop}%
\bibitem [{\citenamefont {Toyoki}\ \emph
  {et~al.}(2015{\natexlab{b}})\citenamefont {Toyoki}, \citenamefont
  {Shiratsuchi}, \citenamefont {Kobane}, \citenamefont {Mitsumata},
  \citenamefont {Kotani}, \citenamefont {Nakamura},\ and\ \citenamefont
  {Nakatani}}]{toyoki_2015-1}%
  \BibitemOpen
  \bibfield  {author} {\bibinfo {author} {\bibfnamefont {K.}~\bibnamefont
  {Toyoki}}, \bibinfo {author} {\bibfnamefont {Y.}~\bibnamefont {Shiratsuchi}},
  \bibinfo {author} {\bibfnamefont {A.}~\bibnamefont {Kobane}}, \bibinfo
  {author} {\bibfnamefont {C.}~\bibnamefont {Mitsumata}}, \bibinfo {author}
  {\bibfnamefont {Y.}~\bibnamefont {Kotani}}, \bibinfo {author} {\bibfnamefont
  {T.}~\bibnamefont {Nakamura}}, \ and\ \bibinfo {author} {\bibfnamefont
  {R.}~\bibnamefont {Nakatani}},\ }\href {\doibase 10.1063/1.4918940}
  {\bibfield  {journal} {\bibinfo  {journal} {Applied Physics Letters}\
  }\textbf {\bibinfo {volume} {106}},\ \bibinfo {pages} {162404} (\bibinfo
  {year} {2015}{\natexlab{b}})}\BibitemShut {NoStop}%
\bibitem [{\citenamefont {Belashchenko}\ \emph {et~al.}(2016)\citenamefont
  {Belashchenko}, \citenamefont {Tchernyshyov}, \citenamefont {Kovalev},\ and\
  \citenamefont {Tretiakov}}]{belashchenko_2016}%
  \BibitemOpen
  \bibfield  {author} {\bibinfo {author} {\bibfnamefont {K.~D.}\ \bibnamefont
  {Belashchenko}}, \bibinfo {author} {\bibfnamefont {O.}~\bibnamefont
  {Tchernyshyov}}, \bibinfo {author} {\bibfnamefont {A.~A.}\ \bibnamefont
  {Kovalev}}, \ and\ \bibinfo {author} {\bibfnamefont {O.~A.}\ \bibnamefont
  {Tretiakov}},\ }\href {\doibase 10.1063/1.4944996} {\bibfield  {journal}
  {\bibinfo  {journal} {Applied Physics Letters}\ }\textbf {\bibinfo {volume}
  {108}},\ \bibinfo {pages} {132403} (\bibinfo {year} {2016})}\BibitemShut
  {NoStop}%
\bibitem [{\citenamefont {Matsuzaki}\ \emph {et~al.}(2014)\citenamefont
  {Matsuzaki}, \citenamefont {Shinohara}, \citenamefont {Shibata},
  \citenamefont {Sahashi},\ and\ \citenamefont {Nozaki}}]{matsuzaki_2014}%
  \BibitemOpen
  \bibfield  {author} {\bibinfo {author} {\bibfnamefont {M.}~\bibnamefont
  {Matsuzaki}}, \bibinfo {author} {\bibfnamefont {K.}~\bibnamefont
  {Shinohara}}, \bibinfo {author} {\bibfnamefont {T.}~\bibnamefont {Shibata}},
  \bibinfo {author} {\bibfnamefont {M.}~\bibnamefont {Sahashi}}, \ and\
  \bibinfo {author} {\bibfnamefont {T.}~\bibnamefont {Nozaki}},\ }\href@noop {}
  {\enquote {\bibinfo {title} {Magnetic recording system and magnetic recording
  device},}\ }\bibinfo {howpublished} {{U.S. Patent No.} 8,724,434} (\bibinfo
  {year} {2014})\BibitemShut {NoStop}%
\bibitem [{\citenamefont {Shibata}\ and\ \citenamefont
  {Sahashi}(2015)}]{shibata_2015}%
  \BibitemOpen
  \bibfield  {author} {\bibinfo {author} {\bibfnamefont {T.}~\bibnamefont
  {Shibata}}\ and\ \bibinfo {author} {\bibfnamefont {M.}~\bibnamefont
  {Sahashi}},\ }\href@noop {} {\enquote {\bibinfo {title} {Magnetization
  controlling element using magnetoelectric effect},}\ }\bibinfo {howpublished}
  {{U.S. Patent App.} 14/532,533} (\bibinfo {year} {2015})\BibitemShut
  {NoStop}%
\bibitem [{\citenamefont {Kota}\ \emph {et~al.}(2013)\citenamefont {Kota},
  \citenamefont {Imamura},\ and\ \citenamefont {Sasaki}}]{kota_2013}%
  \BibitemOpen
  \bibfield  {author} {\bibinfo {author} {\bibfnamefont {Y.}~\bibnamefont
  {Kota}}, \bibinfo {author} {\bibfnamefont {H.}~\bibnamefont {Imamura}}, \
  and\ \bibinfo {author} {\bibfnamefont {M.}~\bibnamefont {Sasaki}},\ }\href
  {\doibase 10.7567/APEX.6.113007} {\bibfield  {journal} {\bibinfo  {journal}
  {Applied Physics Express}\ }\textbf {\bibinfo {volume} {6}},\ \bibinfo
  {pages} {113007} (\bibinfo {year} {2013})}\BibitemShut {NoStop}%
\bibitem [{\citenamefont {Pati}\ \emph {et~al.}(2015)\citenamefont {Pati},
  \citenamefont {Shimomura}, \citenamefont {Nozaki}, \citenamefont {Shibata},\
  and\ \citenamefont {Sahashi}}]{pati_2015}%
  \BibitemOpen
  \bibfield  {author} {\bibinfo {author} {\bibfnamefont {S.~P.}\ \bibnamefont
  {Pati}}, \bibinfo {author} {\bibfnamefont {N.}~\bibnamefont {Shimomura}},
  \bibinfo {author} {\bibfnamefont {T.}~\bibnamefont {Nozaki}}, \bibinfo
  {author} {\bibfnamefont {T.}~\bibnamefont {Shibata}}, \ and\ \bibinfo
  {author} {\bibfnamefont {M.}~\bibnamefont {Sahashi}},\ }\href {\doibase
  10.1063/1.4917263} {\bibfield  {journal} {\bibinfo  {journal} {Journal of
  Applied Physics}\ }\textbf {\bibinfo {volume} {117}},\ \bibinfo {pages}
  {17D137} (\bibinfo {year} {2015})}\BibitemShut {NoStop}%
\bibitem [{\citenamefont {Mu}\ \emph {et~al.}(2013)\citenamefont {Mu},
  \citenamefont {Wysocki},\ and\ \citenamefont {Belashchenko}}]{mu_2013}%
  \BibitemOpen
  \bibfield  {author} {\bibinfo {author} {\bibfnamefont {S.}~\bibnamefont
  {Mu}}, \bibinfo {author} {\bibfnamefont {A.}~\bibnamefont {Wysocki}}, \ and\
  \bibinfo {author} {\bibfnamefont {K.}~\bibnamefont {Belashchenko}},\ }\href
  {\doibase 10.1103/PhysRevB.87.054435} {\bibfield  {journal} {\bibinfo
  {journal} {Physical Review B}\ }\textbf {\bibinfo {volume} {87}},\ \bibinfo
  {pages} {054435} (\bibinfo {year} {2013})}\BibitemShut {NoStop}%
\bibitem [{\citenamefont {Street}\ \emph {et~al.}(2014)\citenamefont {Street},
  \citenamefont {Echtenkamp}, \citenamefont {Komesu}, \citenamefont {Cao},
  \citenamefont {Dowben},\ and\ \citenamefont {Binek}}]{street_2014}%
  \BibitemOpen
  \bibfield  {author} {\bibinfo {author} {\bibfnamefont {M.}~\bibnamefont
  {Street}}, \bibinfo {author} {\bibfnamefont {W.}~\bibnamefont {Echtenkamp}},
  \bibinfo {author} {\bibfnamefont {T.}~\bibnamefont {Komesu}}, \bibinfo
  {author} {\bibfnamefont {S.}~\bibnamefont {Cao}}, \bibinfo {author}
  {\bibfnamefont {P.~A.}\ \bibnamefont {Dowben}}, \ and\ \bibinfo {author}
  {\bibfnamefont {C.}~\bibnamefont {Binek}},\ }\href {\doibase
  10.1063/1.4880938} {\bibfield  {journal} {\bibinfo  {journal} {Applied
  Physics Letters}\ }\textbf {\bibinfo {volume} {104}},\ \bibinfo {pages}
  {222402} (\bibinfo {year} {2014})}\BibitemShut {NoStop}%
\bibitem [{\citenamefont {Carey}\ and\ \citenamefont
  {Berkowitz}(1993)}]{carey_1993}%
  \BibitemOpen
  \bibfield  {author} {\bibinfo {author} {\bibfnamefont {M.~J.}\ \bibnamefont
  {Carey}}\ and\ \bibinfo {author} {\bibfnamefont {A.~E.}\ \bibnamefont
  {Berkowitz}},\ }\href {\doibase 10.1063/1.352426} {\bibfield  {journal}
  {\bibinfo  {journal} {Journal of Applied Physics}\ }\textbf {\bibinfo
  {volume} {73}},\ \bibinfo {pages} {6892} (\bibinfo {year}
  {1993})}\BibitemShut {NoStop}%
\bibitem [{\citenamefont {Ambrose}\ and\ \citenamefont
  {Chien}(1996)}]{ambrose_1996}%
  \BibitemOpen
  \bibfield  {author} {\bibinfo {author} {\bibfnamefont {T.}~\bibnamefont
  {Ambrose}}\ and\ \bibinfo {author} {\bibfnamefont {C.~L.}\ \bibnamefont
  {Chien}},\ }\href {\doibase 10.1103/PhysRevLett.76.1743} {\bibfield
  {journal} {\bibinfo  {journal} {Physical Review Letters}\ }\textbf {\bibinfo
  {volume} {76}},\ \bibinfo {pages} {1743} (\bibinfo {year}
  {1996})}\BibitemShut {NoStop}%
\bibitem [{\citenamefont {van~der Zaag}\ \emph {et~al.}(2000)\citenamefont
  {van~der Zaag}, \citenamefont {Ijiri}, \citenamefont {Borchers},
  \citenamefont {Feiner}, \citenamefont {Wolf}, \citenamefont {Gaines},
  \citenamefont {Erwin},\ and\ \citenamefont {Verheijen}}]{zaag_2000}%
  \BibitemOpen
  \bibfield  {author} {\bibinfo {author} {\bibfnamefont {P.~J.}\ \bibnamefont
  {van~der Zaag}}, \bibinfo {author} {\bibfnamefont {Y.}~\bibnamefont {Ijiri}},
  \bibinfo {author} {\bibfnamefont {J.~A.}\ \bibnamefont {Borchers}}, \bibinfo
  {author} {\bibfnamefont {L.~F.}\ \bibnamefont {Feiner}}, \bibinfo {author}
  {\bibfnamefont {R.~M.}\ \bibnamefont {Wolf}}, \bibinfo {author}
  {\bibfnamefont {J.~M.}\ \bibnamefont {Gaines}}, \bibinfo {author}
  {\bibfnamefont {R.~W.}\ \bibnamefont {Erwin}}, \ and\ \bibinfo {author}
  {\bibfnamefont {M.~A.}\ \bibnamefont {Verheijen}},\ }\href {\doibase
  10.1103/PhysRevLett.84.6102} {\bibfield  {journal} {\bibinfo  {journal}
  {Physical Review Letters}\ }\textbf {\bibinfo {volume} {84}},\ \bibinfo
  {pages} {6102} (\bibinfo {year} {2000})}\BibitemShut {NoStop}%
\bibitem [{\citenamefont {Abarra}\ \emph {et~al.}(1996)\citenamefont {Abarra},
  \citenamefont {Takano}, \citenamefont {Hellman},\ and\ \citenamefont
  {Berkowitz}}]{abarra_1996}%
  \BibitemOpen
  \bibfield  {author} {\bibinfo {author} {\bibfnamefont {E.~N.}\ \bibnamefont
  {Abarra}}, \bibinfo {author} {\bibfnamefont {K.}~\bibnamefont {Takano}},
  \bibinfo {author} {\bibfnamefont {F.}~\bibnamefont {Hellman}}, \ and\
  \bibinfo {author} {\bibfnamefont {A.~E.}\ \bibnamefont {Berkowitz}},\ }\href
  {\doibase 10.1103/PhysRevLett.77.3451} {\bibfield  {journal} {\bibinfo
  {journal} {Physical Review Letters}\ }\textbf {\bibinfo {volume} {77}},\
  \bibinfo {pages} {3451} (\bibinfo {year} {1996})}\BibitemShut {NoStop}%
\bibitem [{\citenamefont {Molina-Ruiz}\ \emph {et~al.}(2011)\citenamefont
  {Molina-Ruiz}, \citenamefont {Lopeand{\'\i}a}, \citenamefont {Pi},
  \citenamefont {Givord}, \citenamefont {Bourgeois},\ and\ \citenamefont
  {Rodr{\'\i}guez-Viejo}}]{molina-ruiz_2011}%
  \BibitemOpen
  \bibfield  {author} {\bibinfo {author} {\bibfnamefont {M.}~\bibnamefont
  {Molina-Ruiz}}, \bibinfo {author} {\bibfnamefont {A.~F.}\ \bibnamefont
  {Lopeand{\'\i}a}}, \bibinfo {author} {\bibfnamefont {F.}~\bibnamefont {Pi}},
  \bibinfo {author} {\bibfnamefont {D.}~\bibnamefont {Givord}}, \bibinfo
  {author} {\bibfnamefont {O.}~\bibnamefont {Bourgeois}}, \ and\ \bibinfo
  {author} {\bibfnamefont {J.}~\bibnamefont {Rodr{\'\i}guez-Viejo}},\ }\href
  {\doibase 10.1103/PhysRevB.83.140407} {\bibfield  {journal} {\bibinfo
  {journal} {Physical Review B}\ }\textbf {\bibinfo {volume} {83}},\ \bibinfo
  {pages} {140407} (\bibinfo {year} {2011})}\BibitemShut {NoStop}%
\bibitem [{\citenamefont {Park}\ \emph {et~al.}(2013)\citenamefont {Park},
  \citenamefont {Jang}, \citenamefont {Kim}, \citenamefont {Park},
  \citenamefont {Koo},\ and\ \citenamefont {Park}}]{park_2013}%
  \BibitemOpen
  \bibfield  {author} {\bibinfo {author} {\bibfnamefont {S.}~\bibnamefont
  {Park}}, \bibinfo {author} {\bibfnamefont {H.}~\bibnamefont {Jang}}, \bibinfo
  {author} {\bibfnamefont {J.-Y.}\ \bibnamefont {Kim}}, \bibinfo {author}
  {\bibfnamefont {B.-G.}\ \bibnamefont {Park}}, \bibinfo {author}
  {\bibfnamefont {T.-Y.}\ \bibnamefont {Koo}}, \ and\ \bibinfo {author}
  {\bibfnamefont {J.-H.}\ \bibnamefont {Park}},\ }\href {\doibase
  10.1209/0295-5075/103/27007} {\bibfield  {journal} {\bibinfo  {journal} {EPL
  (Europhysics Letters)}\ }\textbf {\bibinfo {volume} {103}},\ \bibinfo {pages}
  {27007} (\bibinfo {year} {2013})}\BibitemShut {NoStop}%
\bibitem [{\citenamefont {Frangou}\ \emph {et~al.}(2016)\citenamefont
  {Frangou}, \citenamefont {Oyarzún}, \citenamefont {Auffret}, \citenamefont
  {Vila}, \citenamefont {Gambarelli},\ and\ \citenamefont
  {Baltz}}]{frangou_2016}%
  \BibitemOpen
  \bibfield  {author} {\bibinfo {author} {\bibfnamefont {L.}~\bibnamefont
  {Frangou}}, \bibinfo {author} {\bibfnamefont {S.}~\bibnamefont {Oyarzún}},
  \bibinfo {author} {\bibfnamefont {S.}~\bibnamefont {Auffret}}, \bibinfo
  {author} {\bibfnamefont {L.}~\bibnamefont {Vila}}, \bibinfo {author}
  {\bibfnamefont {S.}~\bibnamefont {Gambarelli}}, \ and\ \bibinfo {author}
  {\bibfnamefont {V.}~\bibnamefont {Baltz}},\ }\href {\doibase
  10.1103/PhysRevLett.116.077203} {\bibfield  {journal} {\bibinfo  {journal}
  {Physical Review Letters}\ }\textbf {\bibinfo {volume} {116}},\ \bibinfo
  {pages} {077203} (\bibinfo {year} {2016})}\BibitemShut {NoStop}%
\bibitem [{\citenamefont {Chambers}\ \emph {et~al.}(2000)\citenamefont
  {Chambers}, \citenamefont {Liang},\ and\ \citenamefont
  {Gao}}]{chambers_2000}%
  \BibitemOpen
  \bibfield  {author} {\bibinfo {author} {\bibfnamefont {S.}~\bibnamefont
  {Chambers}}, \bibinfo {author} {\bibfnamefont {Y.}~\bibnamefont {Liang}}, \
  and\ \bibinfo {author} {\bibfnamefont {Y.}~\bibnamefont {Gao}},\ }\href
  {\doibase 10.1103/PhysRevB.61.13223} {\bibfield  {journal} {\bibinfo
  {journal} {Physical Review B}\ }\textbf {\bibinfo {volume} {61}},\ \bibinfo
  {pages} {13223} (\bibinfo {year} {2000})}\BibitemShut {NoStop}%
\bibitem [{\citenamefont {Shimomura}\ \emph {et~al.}(2015)\citenamefont
  {Shimomura}, \citenamefont {Pati}, \citenamefont {Sato}, \citenamefont
  {Nozaki}, \citenamefont {Shibata}, \citenamefont {Mibu},\ and\ \citenamefont
  {Sahashi}}]{shimomura_2015}%
  \BibitemOpen
  \bibfield  {author} {\bibinfo {author} {\bibfnamefont {N.}~\bibnamefont
  {Shimomura}}, \bibinfo {author} {\bibfnamefont {S.~P.}\ \bibnamefont {Pati}},
  \bibinfo {author} {\bibfnamefont {Y.}~\bibnamefont {Sato}}, \bibinfo {author}
  {\bibfnamefont {T.}~\bibnamefont {Nozaki}}, \bibinfo {author} {\bibfnamefont
  {T.}~\bibnamefont {Shibata}}, \bibinfo {author} {\bibfnamefont
  {K.}~\bibnamefont {Mibu}}, \ and\ \bibinfo {author} {\bibfnamefont
  {M.}~\bibnamefont {Sahashi}},\ }\href {\doibase 10.1063/1.4916304} {\bibfield
   {journal} {\bibinfo  {journal} {Journal of Applied Physics}\ }\textbf
  {\bibinfo {volume} {117}},\ \bibinfo {pages} {17C736} (\bibinfo {year}
  {2015})}\BibitemShut {NoStop}%
\bibitem [{\citenamefont {Mitsui}\ \emph {et~al.}(2016)\citenamefont {Mitsui},
  \citenamefont {Mibu}, \citenamefont {Seto}, \citenamefont {Kurokuzu},
  \citenamefont {Pati}, \citenamefont {Nozaki},\ and\ \citenamefont
  {Sahashi}}]{mitsui_2016}%
  \BibitemOpen
  \bibfield  {author} {\bibinfo {author} {\bibfnamefont {T.}~\bibnamefont
  {Mitsui}}, \bibinfo {author} {\bibfnamefont {K.}~\bibnamefont {Mibu}},
  \bibinfo {author} {\bibfnamefont {M.}~\bibnamefont {Seto}}, \bibinfo {author}
  {\bibfnamefont {M.}~\bibnamefont {Kurokuzu}}, \bibinfo {author}
  {\bibfnamefont {S.~P.}\ \bibnamefont {Pati}}, \bibinfo {author}
  {\bibfnamefont {T.}~\bibnamefont {Nozaki}}, \ and\ \bibinfo {author}
  {\bibfnamefont {M.}~\bibnamefont {Sahashi}},\ }\href {\doibase
  10.7566/JPSJ.85.063601} {\bibfield  {journal} {\bibinfo  {journal} {Journal
  of the Physical Society of Japan}\ }\textbf {\bibinfo {volume} {85}},\
  \bibinfo {pages} {063601} (\bibinfo {year} {2016})}\BibitemShut {NoStop}%
\bibitem [{\citenamefont {Kota}\ \emph
  {et~al.}(2014{\natexlab{b}})\citenamefont {Kota}, \citenamefont {Imamura},\
  and\ \citenamefont {Sasaki}}]{kota_2014}%
  \BibitemOpen
  \bibfield  {author} {\bibinfo {author} {\bibfnamefont {Y.}~\bibnamefont
  {Kota}}, \bibinfo {author} {\bibfnamefont {H.}~\bibnamefont {Imamura}}, \
  and\ \bibinfo {author} {\bibfnamefont {M.}~\bibnamefont {Sasaki}},\ }\href
  {\doibase 10.1109/TMAG.2014.2324014} {\bibfield  {journal} {\bibinfo
  {journal} {IEEE Transactions on Magnetics}\ }\textbf {\bibinfo {volume}
  {50}},\ \bibinfo {pages} {1} (\bibinfo {year}
  {2014}{\natexlab{b}})}\BibitemShut {NoStop}%
\bibitem [{\citenamefont {Shimomura}\ \emph {et~al.}(2016)\citenamefont
  {Shimomura}, \citenamefont {Pati}, \citenamefont {Nozaki}, \citenamefont
  {Shibata},\ and\ \citenamefont {Sahashi}}]{shimomura_2016}%
  \BibitemOpen
  \bibfield  {author} {\bibinfo {author} {\bibfnamefont {N.}~\bibnamefont
  {Shimomura}}, \bibinfo {author} {\bibfnamefont {S.~P.}\ \bibnamefont {Pati}},
  \bibinfo {author} {\bibfnamefont {T.}~\bibnamefont {Nozaki}}, \bibinfo
  {author} {\bibfnamefont {T.}~\bibnamefont {Shibata}}, \ and\ \bibinfo
  {author} {\bibfnamefont {M.}~\bibnamefont {Sahashi}},\ }\href
  {http://arxiv.org/abs/1605.03680} {\bibfield  {journal} {\bibinfo  {journal}
  {arXiv:1605.03680 [cond-mat]}\ } (\bibinfo {year} {2016})}\BibitemShut
  {NoStop}%
\bibitem [{\citenamefont {Evans}\ \emph {et~al.}(2014)\citenamefont {Evans},
  \citenamefont {Fan}, \citenamefont {Chureemart}, \citenamefont {Ostler},
  \citenamefont {Ellis},\ and\ \citenamefont {Chantrell}}]{evans_2014}%
  \BibitemOpen
  \bibfield  {author} {\bibinfo {author} {\bibfnamefont {R.~F.~L.}\
  \bibnamefont {Evans}}, \bibinfo {author} {\bibfnamefont {W.~J.}\ \bibnamefont
  {Fan}}, \bibinfo {author} {\bibfnamefont {P.}~\bibnamefont {Chureemart}},
  \bibinfo {author} {\bibfnamefont {T.~A.}\ \bibnamefont {Ostler}}, \bibinfo
  {author} {\bibfnamefont {M.~O.~A.}\ \bibnamefont {Ellis}}, \ and\ \bibinfo
  {author} {\bibfnamefont {R.~W.}\ \bibnamefont {Chantrell}},\ }\href {\doibase
  10.1088/0953-8984/26/10/103202} {\bibfield  {journal} {\bibinfo  {journal}
  {Journal of Physics: Condensed Matter}\ }\textbf {\bibinfo {volume} {26}},\
  \bibinfo {pages} {103202} (\bibinfo {year} {2014})}\BibitemShut {NoStop}%
\bibitem [{\citenamefont {Belashchenko}(2010)}]{belashchenko_2010}%
  \BibitemOpen
  \bibfield  {author} {\bibinfo {author} {\bibfnamefont {K.~D.}\ \bibnamefont
  {Belashchenko}},\ }\href {\doibase 10.1103/PhysRevLett.105.147204} {\bibfield
   {journal} {\bibinfo  {journal} {Physical Review Letters}\ }\textbf {\bibinfo
  {volume} {105}},\ \bibinfo {pages} {147204} (\bibinfo {year}
  {2010})}\BibitemShut {NoStop}%
\bibitem [{\citenamefont {McGuire}\ \emph {et~al.}(1956)\citenamefont
  {McGuire}, \citenamefont {Scott},\ and\ \citenamefont
  {Grannis}}]{mcguire_1956}%
  \BibitemOpen
  \bibfield  {author} {\bibinfo {author} {\bibfnamefont {T.~R.}\ \bibnamefont
  {McGuire}}, \bibinfo {author} {\bibfnamefont {E.~J.}\ \bibnamefont {Scott}},
  \ and\ \bibinfo {author} {\bibfnamefont {F.~H.}\ \bibnamefont {Grannis}},\
  }\href {\doibase 10.1103/PhysRev.102.1000} {\bibfield  {journal} {\bibinfo
  {journal} {Physical Review}\ }\textbf {\bibinfo {volume} {102}},\ \bibinfo
  {pages} {1000} (\bibinfo {year} {1956})}\BibitemShut {NoStop}%
\bibitem [{\citenamefont {Iyama}\ and\ \citenamefont
  {Kimura}(2013)}]{iyama_2013}%
  \BibitemOpen
  \bibfield  {author} {\bibinfo {author} {\bibfnamefont {A.}~\bibnamefont
  {Iyama}}\ and\ \bibinfo {author} {\bibfnamefont {T.}~\bibnamefont {Kimura}},\
  }\href {\doibase 10.1103/PhysRevB.87.180408} {\bibfield  {journal} {\bibinfo
  {journal} {Physical Review B}\ }\textbf {\bibinfo {volume} {87}},\ \bibinfo
  {pages} {180408} (\bibinfo {year} {2013})}\BibitemShut {NoStop}%
\bibitem [{\citenamefont {Shiratsuchi}\ \emph {et~al.}(2015)\citenamefont
  {Shiratsuchi}, \citenamefont {Kotani}, \citenamefont {Yoshida}, \citenamefont
  {Yoshikawa}, \citenamefont {Toyoki}, \citenamefont {Kobane}, \citenamefont
  {Nakatani},\ and\ \citenamefont {Nakamura}}]{shiratsuchi_2015}%
  \BibitemOpen
  \bibfield  {author} {\bibinfo {author} {\bibfnamefont {Y.}~\bibnamefont
  {Shiratsuchi}}, \bibinfo {author} {\bibfnamefont {Y.}~\bibnamefont {Kotani}},
  \bibinfo {author} {\bibfnamefont {S.}~\bibnamefont {Yoshida}}, \bibinfo
  {author} {\bibfnamefont {Y.}~\bibnamefont {Yoshikawa}}, \bibinfo {author}
  {\bibfnamefont {K.}~\bibnamefont {Toyoki}}, \bibinfo {author} {\bibfnamefont
  {A.}~\bibnamefont {Kobane}}, \bibinfo {author} {\bibfnamefont
  {R.}~\bibnamefont {Nakatani}}, \ and\ \bibinfo {author} {\bibfnamefont
  {T.}~\bibnamefont {Nakamura}},\ }\href {\doibase 10.3934/matersci.2015.4.484}
  {\bibfield  {journal} {\bibinfo  {journal} {AIMS Materials Science}\ }\textbf
  {\bibinfo {volume} {2}},\ \bibinfo {pages} {484} (\bibinfo {year}
  {2015})}\BibitemShut {NoStop}%
\bibitem [{\citenamefont {Barber}(1983)}]{barber_1983}%
  \BibitemOpen
  \bibfield  {author} {\bibinfo {author} {\bibfnamefont {M.}~\bibnamefont
  {Barber}},\ }in\ \href@noop {} {\emph {\bibinfo {booktitle} {Phase
  transitions and critical phenomena}}},\ Vol.~\bibinfo {volume} {8},\ \bibinfo
  {editor} {edited by\ \bibinfo {editor} {\bibfnamefont {M.~S.}\ \bibnamefont
  {Green}}\ and\ \bibinfo {editor} {\bibfnamefont {J.~L.}\ \bibnamefont
  {Lebowitz}}}\ (\bibinfo  {publisher} {{Academic Press}},\ \bibinfo {year}
  {1983})\ pp.\ \bibinfo {pages} {145--266}\BibitemShut {NoStop}%
\bibitem [{\citenamefont {Privman}(1990)}]{privman_1990}%
  \BibitemOpen
  \bibfield  {author} {\bibinfo {author} {\bibfnamefont {V.}~\bibnamefont
  {Privman}},\ }\href@noop {} {\emph {\bibinfo {title} {Finite {{Size Scaling}}
  and {{Numerical Simulation}} of {{Statistical Systems}}}}}\ (\bibinfo
  {publisher} {{WORLD SCIENTIFIC}},\ \bibinfo {year} {1990})\BibitemShut
  {NoStop}%
\bibitem [{\citenamefont {Fisher}\ and\ \citenamefont
  {Barber}(1972)}]{fisher_1972}%
  \BibitemOpen
  \bibfield  {author} {\bibinfo {author} {\bibfnamefont {M.~E.}\ \bibnamefont
  {Fisher}}\ and\ \bibinfo {author} {\bibfnamefont {M.~N.}\ \bibnamefont
  {Barber}},\ }\href {\doibase 10.1103/PhysRevLett.28.1516} {\bibfield
  {journal} {\bibinfo  {journal} {Physical Review Letters}\ }\textbf {\bibinfo
  {volume} {28}},\ \bibinfo {pages} {1516} (\bibinfo {year}
  {1972})}\BibitemShut {NoStop}%
\bibitem [{\citenamefont {Binder}\ and\ \citenamefont
  {Hohenberg}(1974)}]{binder_1974}%
  \BibitemOpen
  \bibfield  {author} {\bibinfo {author} {\bibfnamefont {K.}~\bibnamefont
  {Binder}}\ and\ \bibinfo {author} {\bibfnamefont {P.~C.}\ \bibnamefont
  {Hohenberg}},\ }\href {\doibase 10.1103/PhysRevB.9.2194} {\bibfield
  {journal} {\bibinfo  {journal} {Physical Review B}\ }\textbf {\bibinfo
  {volume} {9}},\ \bibinfo {pages} {2194} (\bibinfo {year} {1974})}\BibitemShut
  {NoStop}%
\bibitem [{\citenamefont {Foner}(1963)}]{foner_1963}%
  \BibitemOpen
  \bibfield  {author} {\bibinfo {author} {\bibfnamefont {S.}~\bibnamefont
  {Foner}},\ }\href {\doibase 10.1103/PhysRev.130.183} {\bibfield  {journal}
  {\bibinfo  {journal} {Physical Review}\ }\textbf {\bibinfo {volume} {130}},\
  \bibinfo {pages} {183} (\bibinfo {year} {1963})}\BibitemShut {NoStop}%
\bibitem [{\citenamefont {Samuelsen}\ \emph {et~al.}(1970)\citenamefont
  {Samuelsen}, \citenamefont {Hutchings},\ and\ \citenamefont
  {Shirane}}]{samuelsen_1970}%
  \BibitemOpen
  \bibfield  {author} {\bibinfo {author} {\bibfnamefont {E.~J.}\ \bibnamefont
  {Samuelsen}}, \bibinfo {author} {\bibfnamefont {M.~T.}\ \bibnamefont
  {Hutchings}}, \ and\ \bibinfo {author} {\bibfnamefont {G.}~\bibnamefont
  {Shirane}},\ }\href {\doibase 10.1016/0031-8914(70)90158-8} {\bibfield
  {journal} {\bibinfo  {journal} {Physica}\ }\textbf {\bibinfo {volume} {48}},\
  \bibinfo {pages} {13} (\bibinfo {year} {1970})}\BibitemShut {NoStop}%
\bibitem [{\citenamefont {Finger}\ and\ \citenamefont
  {Hazen}(1980)}]{finger_1980}%
  \BibitemOpen
  \bibfield  {author} {\bibinfo {author} {\bibfnamefont {L.~W.}\ \bibnamefont
  {Finger}}\ and\ \bibinfo {author} {\bibfnamefont {R.~M.}\ \bibnamefont
  {Hazen}},\ }\href {\doibase 10.1063/1.327451} {\bibfield  {journal} {\bibinfo
   {journal} {Journal of Applied Physics}\ }\textbf {\bibinfo {volume} {51}},\
  \bibinfo {pages} {5362} (\bibinfo {year} {1980})}\BibitemShut {NoStop}%
\bibitem [{\citenamefont {Brown}\ \emph {et~al.}(2002)\citenamefont {Brown},
  \citenamefont {Forsyth}, \citenamefont {Leli{\`e}vre-Berna},\ and\
  \citenamefont {Tasset}}]{brown_2002}%
  \BibitemOpen
  \bibfield  {author} {\bibinfo {author} {\bibfnamefont {P.~J.}\ \bibnamefont
  {Brown}}, \bibinfo {author} {\bibfnamefont {J.~B.}\ \bibnamefont {Forsyth}},
  \bibinfo {author} {\bibfnamefont {E.}~\bibnamefont {Leli{\`e}vre-Berna}}, \
  and\ \bibinfo {author} {\bibfnamefont {F.}~\bibnamefont {Tasset}},\ }\href
  {\doibase 10.1088/0953-8984/14/8/323} {\bibfield  {journal} {\bibinfo
  {journal} {Journal of Physics: Condensed Matter}\ }\textbf {\bibinfo {volume}
  {14}},\ \bibinfo {pages} {1957} (\bibinfo {year} {2002})}\BibitemShut
  {NoStop}%
\bibitem [{\citenamefont {Tang}\ \emph {et~al.}(2003)\citenamefont {Tang},
  \citenamefont {Smith}, \citenamefont {Zink}, \citenamefont {Hellman},\ and\
  \citenamefont {Berkowitz}}]{tang_2003}%
  \BibitemOpen
  \bibfield  {author} {\bibinfo {author} {\bibfnamefont {Y.~J.}\ \bibnamefont
  {Tang}}, \bibinfo {author} {\bibfnamefont {D.~J.}\ \bibnamefont {Smith}},
  \bibinfo {author} {\bibfnamefont {B.~L.}\ \bibnamefont {Zink}}, \bibinfo
  {author} {\bibfnamefont {F.}~\bibnamefont {Hellman}}, \ and\ \bibinfo
  {author} {\bibfnamefont {A.~E.}\ \bibnamefont {Berkowitz}},\ }\href {\doibase
  10.1103/PhysRevB.67.054408} {\bibfield  {journal} {\bibinfo  {journal}
  {Physical Review B}\ }\textbf {\bibinfo {volume} {67}},\ \bibinfo {pages}
  {054408} (\bibinfo {year} {2003})}\BibitemShut {NoStop}%
\bibitem [{\citenamefont {Wang}\ \emph {et~al.}(2000)\citenamefont {Wang},
  \citenamefont {Chaka},\ and\ \citenamefont {Scheffler}}]{wang_2000}%
  \BibitemOpen
  \bibfield  {author} {\bibinfo {author} {\bibfnamefont {X.-G.}\ \bibnamefont
  {Wang}}, \bibinfo {author} {\bibfnamefont {A.}~\bibnamefont {Chaka}}, \ and\
  \bibinfo {author} {\bibfnamefont {M.}~\bibnamefont {Scheffler}},\ }\href
  {\doibase 10.1103/PhysRevLett.84.3650} {\bibfield  {journal} {\bibinfo
  {journal} {Physical Review Letters}\ }\textbf {\bibinfo {volume} {84}},\
  \bibinfo {pages} {3650} (\bibinfo {year} {2000})}\BibitemShut {NoStop}%
\end{thebibliography}%

\end{document}